\documentclass[%
  reprint,
  superscriptaddress,
  longbibliography,
  preprintnumbers,
  amsmath,
  amssymb,
  aps,
  prl,
  floatfix,
  tightenlines
]{revtex4-2}

\usepackage{physics}
\usepackage{bm}

\usepackage{changes}
\usepackage[colorlinks=true,
citecolor=green,
linkcolor=purple,
anchorcolor=black,
urlcolor=purple]{hyperref}

\begin{document}

% \preprint{APS/123-QED}

\title{Dissipation-Selected Resonant Fronts in a Driven-Dissipative Bose-Hubbard Lattice}

\author{Wei-Guo Ma}
\email{weiguo.m@iphy.ac.cn}
\affiliation{Beijing National Laboratory for Condensed Matter Physics,
Institute of Physics, Chinese Academy of Sciences, Beijing 100190, China}
\affiliation{School of Physical Sciences, University of Chinese Academy of Sciences, Beijing 100049, China}

\author{Heng Fan}
\email{hfan@iphy.ac.cn}
\affiliation{Beijing National Laboratory for Condensed Matter Physics,
Institute of Physics, Chinese Academy of Sciences, Beijing 100190, China}
\affiliation{School of Physical Sciences, University of Chinese Academy of Sciences, Beijing 100049, China}
\affiliation{Beijing Key Laboratory of Advanced Quantum Technology,
Beijing Academy of Quantum Information Sciences, Beijing 100193, China}
\affiliation{Hefei National Laboratory, Hefei 230088, China}
\affiliation{Songshan Lake Materials Laboratory, Dongguan, Guangdong 523808, China}

\date{\today}

\begin{abstract}
Spatially structured dissipation organizes driven quantum matter beyond Hamiltonian control. We show that a dissipation gradient combined with a Stark-induced detuning ramp selects a nonlinear resonance slice in a two-dimensional driven-dissipative Bose-Hubbard lattice, producing a pinned density front in generalized Gross-Pitaevskii simulations. The underlying resonance condition fixes the front position, while its Airy-like profile obeys a width scaling set by tunneling stiffness and the effective detuning slope. Treating the front as an emergent interface explains how tuning the selected resonance toward the minimum-loss side yields Peierls-Nabarro depinning steps, discrete transverse pattern locking, spatiotemporal chaos, and minimum-loss localization. Center-of-mass and generalized-imbalance diagnostics map these outcomes into a dynamical phase diagram as detuning-ramp slope and dissipation-gradient strength vary. The results suggest structured dissipation as a mechanism for reconfigurable transport barriers and nonequilibrium interfaces in programmable bosonic lattices.
\end{abstract}

\maketitle

\textit{Introduction.}---Driven-dissipative quantum many-body systems~\cite{Sieberer_2016, Noh_2017, Dutta_2015} provide a fertile platform for exploring non-equilibrium physics, where the intrinsic competition between coherent driving and dissipation gives rise to dynamical regimes and steady states that transcend traditional frameworks~\cite{PhysRevX.7.011016,PhysRevX.7.011012,PhysRevA.98.042118,PhysRevLett.118.247402,PhysRevX.7.021020,aar4003,Owen_2018,PhysRevB.95.134431}. Within this landscape, the driven Bose-Hubbard model has served as a paradigmatic testbed~\cite{PhysRev.129.959,PhysRevB.40.546,Kordas:2015aa,PhysRevResearch.6.L032067,PhysRevA.81.061801,Jouanny:2025aa, PhysRevA.84.043827,PhysRevX.7.011016,PhysRevLett.126.180503,PhysRevLett.124.160604,PhysRevB.87.085131}, with phenomena ranging from dissipative phase transitions and non-Hermitian criticality~\cite{PhysRevA.97.013853, PhysRevLett.130.063601, MIYAZAKI200341, ceulemans2023nonequilibrium, PhysRevResearch.6.L032067, PhysRevA.95.012128, PhysRevA.96.023839} to bistability, tunneling-induced instabilities~\cite{PhysRevA.90.063821, PhysRevA.95.043833, Rodriguez:2016aa, PhysRevLett.110.233601}, and two-dimensional pattern formation~\cite{PhysRevLett.125.115301, zhang2020pattern}. 

Across diverse quantum architectures, experimental realizations of driven lattices have achieved a level of exquisite tunability, where techniques such as Floquet engineering and lattice shaking grant precise access to the parameter space~\cite{RevModPhys.89.011004,PhysRevLett.116.205301, wangzt2025,Leghtas2015,Kannan:2020aa,PhysRevA.103.023710, PhysRevLett.120.140404, PhysRevLett.126.043602}. Propelling this frontier further, recent experimental breakthroughs in circuit quantum electrodynamics, programmable
atomic lattices, and polaritonic microcavities have unlocked versatile spatial control over system parameters~\cite{science.1074464,science.1140990,PhysRevX.7.011016,Hartmann2016,Ebadi:2021aa,Browaeys:2020aa,RevModPhys.93.025005,Ma:2019aa,Klembt:2018aa,Carusotto:2020aa,RevModPhys.93.025001,GrossBloch2017,PhysRevLett.110.035302}. These capabilities enable the engineering of spatially inhomogeneous driving or dissipation profiles, opening avenues to explore transport effects distinct from those in homogeneous settings~\cite{ceulemans2023nonequilibrium,kosior2024nonequilibrium,PhysRevLett.110.035302,PhysRevLett.115.050601, PhysRevLett.128.093601}. We use this freedom to place a Stark-induced detuning ramp against a co-linear dissipation gradient. The resulting detuning-dissipation imbalance provides a minimal setting for asking how a driven quantum fluid organizes under spatially competing detuning and dissipation.

In this Letter, we investigate the nonequilibrium dynamics in a two-dimensional driven-dissipative Bose-Hubbard lattice using generalized Gross-Pitaevskii simulations within the truncated-Wigner approach. We find that a dissipation gradient combined with a Stark-induced detuning ramp produces a density wave with a sharp front. The front is pinned by a nonlinear resonance condition that fixes its position, and its Airy-like width follows from the balance between tunneling stiffness and the effective detuning slope. Tuning the ramp moves this selected front and drives a sequence of long-time regimes: pinned, spatiotemporal chaos, and localization at the minimum-loss side. Within the pinned regime, the front can remain uniform or develop discrete transverse pattern locking, and lattice discreteness produces a macroscopic particle-number staircase consistent with Peierls-Nabarro depinning. We then use center-of-mass and generalized-imbalance diagnostics to assemble these outcomes into a dynamical phase diagram of a dissipation-selected front. More broadly, our results uncover a resonance mechanism for front selection in open quantum lattices and suggest a route to engineer tunable barriers and reconfigurable nonequilibrium interfaces in platforms including photonic and polaritonic arrays and superconducting-circuit lattices.

\begin{figure}[t]
  \centering
  \includegraphics[width=\linewidth]{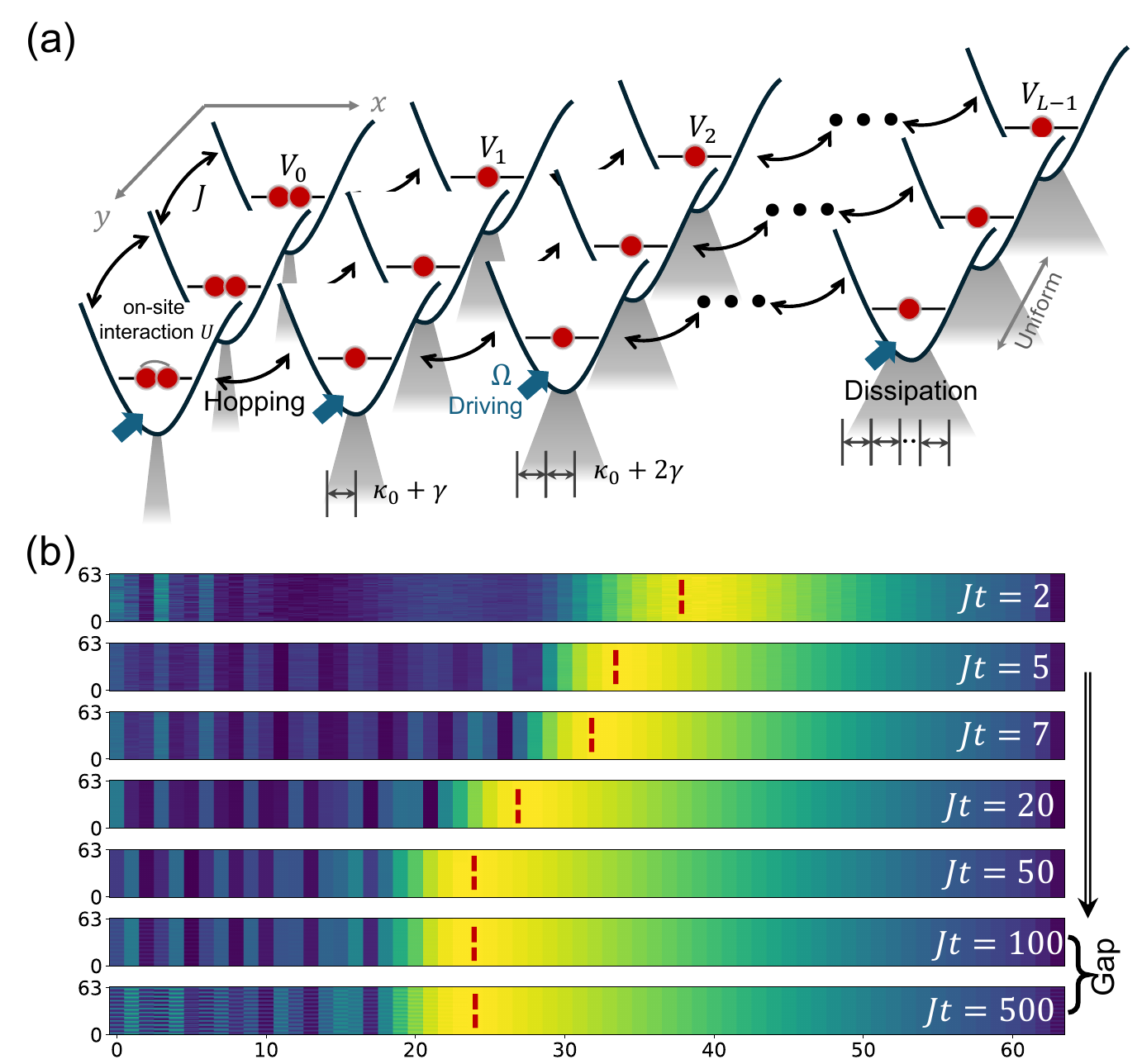}
  \caption{Setup and density front formation in a two-dimensional driven-dissipative Bose-Hubbard lattice. (a) Schematic of the model. The system consists of an array of nonlinear resonators with on-site interaction $U$, driven by a single-photon field $\Omega$ at frequency $\omega_d$. While the system remains uniform along the $y$-axis, it varies along the $x$-axis. Specifically, the linear Stark potential $Fx_j$ induces the detuning ramp, while the dissipation gradient $\kappa_j = \kappa_0 + \gamma x_j$ sets the spatially graded loss rate. Bosons hop between nearest neighbors at rate $J$. (b) Time evolution of the condensate density, showing the formation and pinning of the front (red dashed line). Brighter colors indicate higher local occupation.}
  \label{fig:model_time_evolution}
\end{figure}

\textit{Model and its Hamiltonian.}---We consider a two-dimensional driven-dissipative Bose-Hubbard lattice incorporating a Stark-induced detuning ramp and a spatially graded loss rate, as illustrated in Fig.~\!\ref{fig:model_time_evolution}(a). The unitary dynamics are governed by
\begin{equation}
\begin{aligned}
\hat{H} &= \sum_{j}(\omega_{0} + F x_j)\hat{a}_j^\dagger \hat{a}_j + \frac{U}{2}\sum_{j}\hat{a}_j^\dagger \hat{a}_j^\dagger \hat{a}_j \hat{a}_j \\
&- J\sum_{\langle j,k \rangle}(\hat{a}_j^\dagger \hat{a}_k + \hat{a}_k^\dagger \hat{a}_j) + \Omega\sum_{j}\left(\hat{a}_j e^{i\omega_d t} + \hat{a}_j^\dagger e^{-i\omega_d t}\right),
\label{eq:hamiltonian_main}
\end{aligned}
\end{equation}
where $\hat{a}_j$ is the annihilation operator at site $j$. The parameters $U$, $J$, and $\Omega$ denote the on-site interaction, nearest-neighbor hopping, and drive strength (at frequency $\omega_d$), respectively. The coefficient $F \geq 0$ sets a linear Stark potential along the $+x$ direction and therefore the detuning ramp. Moving to a frame rotating at $\omega_d$ eliminates the explicit time dependence, yielding the effective Hamiltonian $\hat{H}_{\text{rot}}=\sum_{j}(\Delta\omega + Fx_j)\hat{a}_j^\dagger \hat{a}_j + \frac{1}{2}U\sum_j\hat{a}_j^\dagger \hat{a}_j^\dagger \hat{a}_j \hat{a}_j - J\sum_{\langle j,k\rangle}(\hat{a}_j^\dagger \hat{a}_k + \hat{a}_k^\dagger \hat{a}_j) + \Omega\sum_{j}(\hat{a}_j + \hat{a}_j^\dagger)$, with $\Delta\omega = \omega_0 - \omega_d$. The open system dynamics obey the master equation $\partial_t \rho = -i[\hat{H}_{\text{rot}}, \rho] + \frac{1}{2}\sum_j \kappa_j (2\hat{a}_j\rho \hat{a}_j^\dagger - \hat{a}_j^\dagger \hat{a}_j\rho - \rho \hat{a}_j^\dagger \hat{a}_j)$, where the loss rate follows $\kappa_j = \kappa_0 + \gamma x_j$, with $\kappa_0$ the uniform baseline loss rate and $\gamma \geq 0$ the dissipation gradient. To access large-scale dynamics, we use $\alpha_j = \langle \hat{a}_j \rangle$, which gives the generalized Gross-Pitaevskii equation (gGPE):
\begin{equation}
\begin{aligned}
i\frac{\dd\alpha_j}{\dd t} =&\left[\Delta\omega + Fx_j - \frac{i}{2}(\kappa_0 + \gamma x_j)\right]\alpha_j \\
&\quad + U|\alpha_j|^2\alpha_j - J\sum_{k\in\langle j\rangle}\alpha_k + \Omega.
\label{eq:GrossPitaevskiiTilt}
\end{aligned}
\end{equation}
The drive and dissipation therefore compete not only onsite but also through a spatial detuning-dissipation imbalance. To include fluctuation effects that can seed instabilities beyond mean-field dynamics, we employ the initial-value truncated Wigner approximation (IV-TWA)~\cite{Alice_Sinatra_2002, Carusotto2013, PhysRevA.106.042406, PhysRevA.99.043627} and evolve each stochastic realization according to Eq.~\!\eqref{eq:GrossPitaevskiiTilt}. Quantum fluctuations enter through stochastic sampling of the initial Wigner distribution, and observables are obtained from ensemble averages over many trajectories~\cite{supplementary}.
\begin{figure*}[t]
  \centering
  \includegraphics[width=0.9\linewidth]{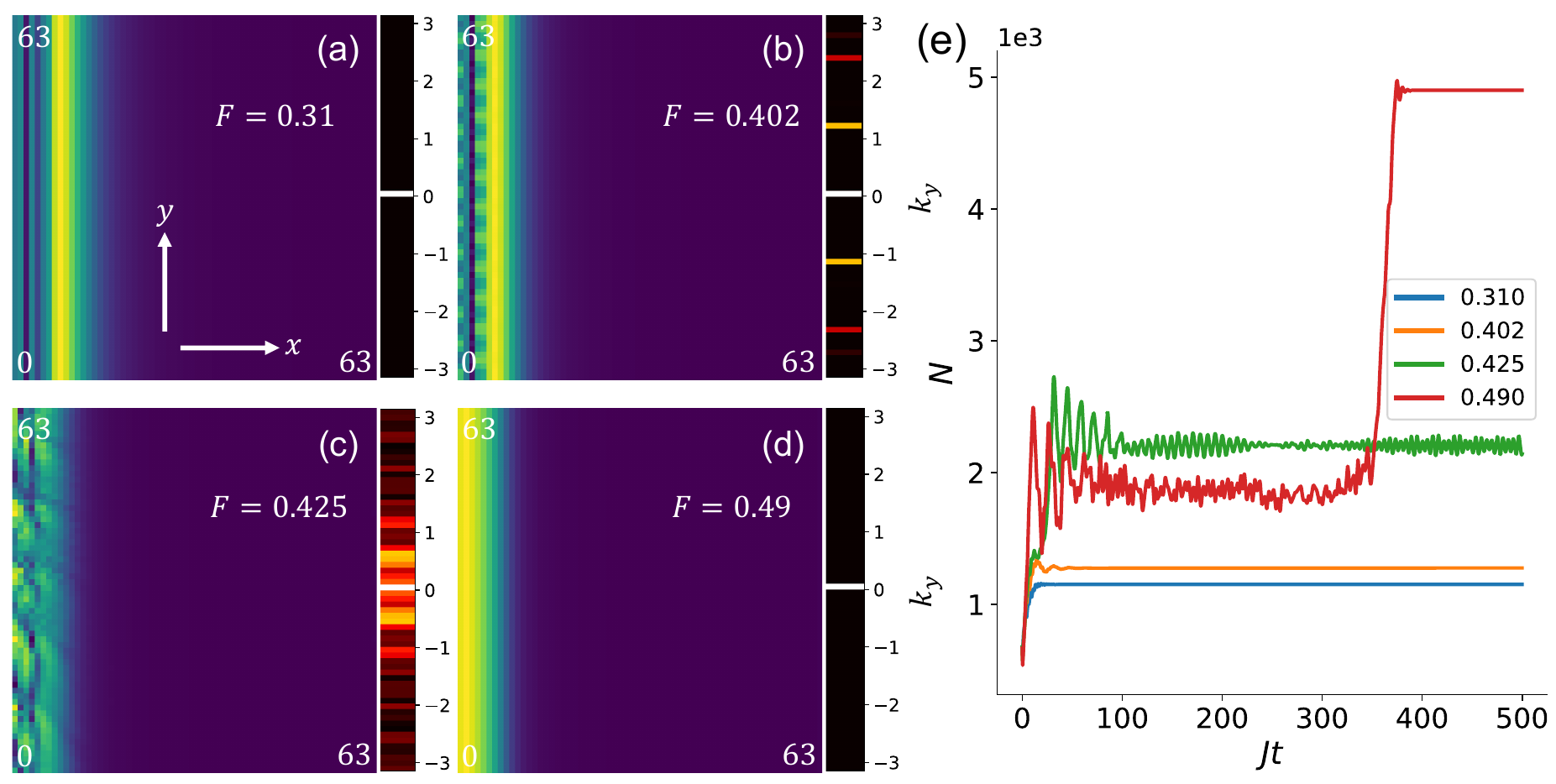}
  \caption{Long-time density profiles and transverse spectra at $Jt=500$ for fixed dissipation gradient $\gamma = 0.133$ and baseline loss rate $\kappa_0=0.05$, with the detuning-ramp slope $F$ set by the Stark tilt. (a) A uniform pinned front at $F=0.31$, with a dominant $k_y=0$ spectral component. (b) A pattern-locked front at $F=0.402$, with discrete transverse harmonics. (c) A spatiotemporally chaotic state at $F = 0.425$, with broadband spectral mixing. (d) A minimum-loss-localized state at $F=0.49$, again dominated by $k_y=0$. (e) Corresponding time evolution of the total particle number $N$ for various $F$.}
  \label{fig:dynamics_time}
\end{figure*}

\textit{Dissipation-selected front.}---The density front is selected by a local nonlinear resonance. We perform numerical simulations on a $64\times 64$ lattice with periodic hopping boundaries; the coordinate-dependent detuning and dissipation profiles are imposed along $x$, with $x=0$ denoting the minimum-loss side. Since the dynamics is spatially resolved, we use $\alpha_j=\alpha_{x,y}$, where $(x,y)$ is the two-dimensional lattice coordinate. Unless otherwise stated, the system parameters are set to $U{=}0.2$, $\Omega=1$, $J=1$, $\Delta\omega=0.6$, and $\kappa_0=0.05$. The system is evolved according to Eq.~\!\eqref{eq:GrossPitaevskiiTilt} up to $Jt=500$, which satisfies the long-time criterion $t \gg \kappa_0^{-1}$. 

Figure~\!\ref{fig:model_time_evolution}(b) shows a density wave with a sharp front that separates regions of distinct density and becomes pinned. Because the pinned front is nearly uniform along the transverse direction $y$, we approximate $\alpha_{x,y}\approx \alpha_x$ and absorb the transverse hopping into a constant detuning shift. This reduction yields the effective one-dimensional steady-state equation
$-J(\alpha_{x+1}+\alpha_{x-1}-2\alpha_x)+[\Delta_{1\mathrm{D}}(x)-\frac{i}{2}(\kappa_0 + \gamma x)]\alpha_x+U|\alpha_x|^2\alpha_x+\Omega=0$, where $\Delta_{1\mathrm{D}}(x)=\Delta\omega+Fx-4J$ incorporates the transverse coordination shift as an offset. Within a local-density approximation that neglects the discrete curvature term $-J(\alpha_{x+1}+\alpha_{x-1}-2\alpha_x)$, each $x$ slice is an isolated Kerr oscillator~\cite{PhysRevLett.58.2209}. The local density $N(x)=|\alpha_x|^2$ thus follows the input-output relation $\Omega^2 = N(x)[\Delta_{\text{eff}}^2(x)+\tfrac14(\kappa_0 + \gamma x)^2]$, with $\Delta_{\text{eff}}(x)=\Delta_{1\mathrm{D}}(x)-U N(x)$. The front is pinned at the resonant slice $x_s$ such that $\Delta_{\text{eff}}(x_s)=0$. Since both the detuning ramp and the loss rate increase monotonically with $x$, the resonance condition admits a unique solution for $x_s$ and does not rely on local bistability. 

The same resonance picture predicts the front width once the curvature term is restored. Defining the effective gradient steepness as $S \equiv \partial_x\Delta_{\text{eff}}|_{x_s}$, we linearize $\Delta_{\text{eff}}(x)\approx S(x-x_s)$, which yields an Airy-type envelope whose characteristic length defines the front width $w$. The resulting width is set by the competition between tunneling stiffness and the effective detuning slope and obeys the scaling
\begin{equation}
w \propto \left(\frac{J}{S}\right)^{1/3},\; \text{with}\;\, S\approx F + \frac{8U\gamma\Omega^2}{\big(\kappa_0 + \gamma x_s\big)^3}.
\end{equation} 
For derivation details, including the downstream decay length and an additional detuning correction that refines the resonance pinning condition, see~\cite{supplementary}.

\textit{Front dynamics.}---Once selected, the front behaves as an interface whose dynamics are controlled by the resonant slice. We vary the detuning-ramp slope $F$ at fixed $\gamma=0.133$ and compare the long-time density profiles and transverse spectra in Figs.~\!\ref{fig:dynamics_time}(a-d), together with the total particle number $N(t)=\sum_j|\alpha_j(t)|^2$ in Fig.~\!\ref{fig:dynamics_time}(e). In the shallow-ramp regime ($F{=}0.31$), the selected front is pinned and nearly uniform, with a single dominant $k_y=0$ spectral component. Increasing $F$ to $0.402$ shifts the selected front toward the minimum-loss side, where transverse modulation instability drives it into a pattern-locked configuration. We describe the front dynamics via the ansatz $\alpha \approx \alpha_0(x-X)$, with the displacement $X(y,t)$ evolving according to the amplitude equation
\begin{equation}
\partial_t X = - D\partial_y^2 X - K\partial_y^4 X + \Lambda(\partial_y X)^2 + \cdots,
\label{eq:amp_main_final}
\end{equation}
where $D$, $K$, and $\Lambda$ represent the effective tension, stiffness, and nonlinearity, respectively. Linearizing Eq.~\!\eqref{eq:amp_main_final} yields the growth rate $\sigma(k_y) = -D k_y^2 - K k_y^4$. With $D<0$ and $K>0$, a band of long-wavelength modes becomes unstable, selecting discrete harmonics on the finite lattice, as seen in the transverse spectrum of Fig.~\!\ref{fig:dynamics_time}(b) (see details in~\cite{supplementary}). At $F=0.425$, the same front enters a regime of spatiotemporal chaos. Equation~\!\eqref{eq:amp_main_final} is recognized as the canonical Kuramoto-Sivashinsky model~\cite{PhysRevA.90.023615, PhysRevB.91.045301, RevModPhys.97.025004}, where the nonlinearity $\Lambda(\partial_y X)^2$ transfers energy from unstable low-$k$ modes to dissipative high-$k$ modes, sustaining a persistent turbulent state characterized by the broadband spectral mixing in Fig.~\!\ref{fig:dynamics_time}(c). At a stronger ramp ($F=0.49$), the selected front localizes at the minimum-loss side and reverts to a minimum-loss-localized profile. The time-resolved Bogoliubov-de Gennes (BdG) analysis in~\cite{supplementary} tracks when $\lambda_{\text{max}}$ becomes positive around the evolving state.
\begin{figure}[htbp]
  \centering
  \includegraphics[width=\linewidth]{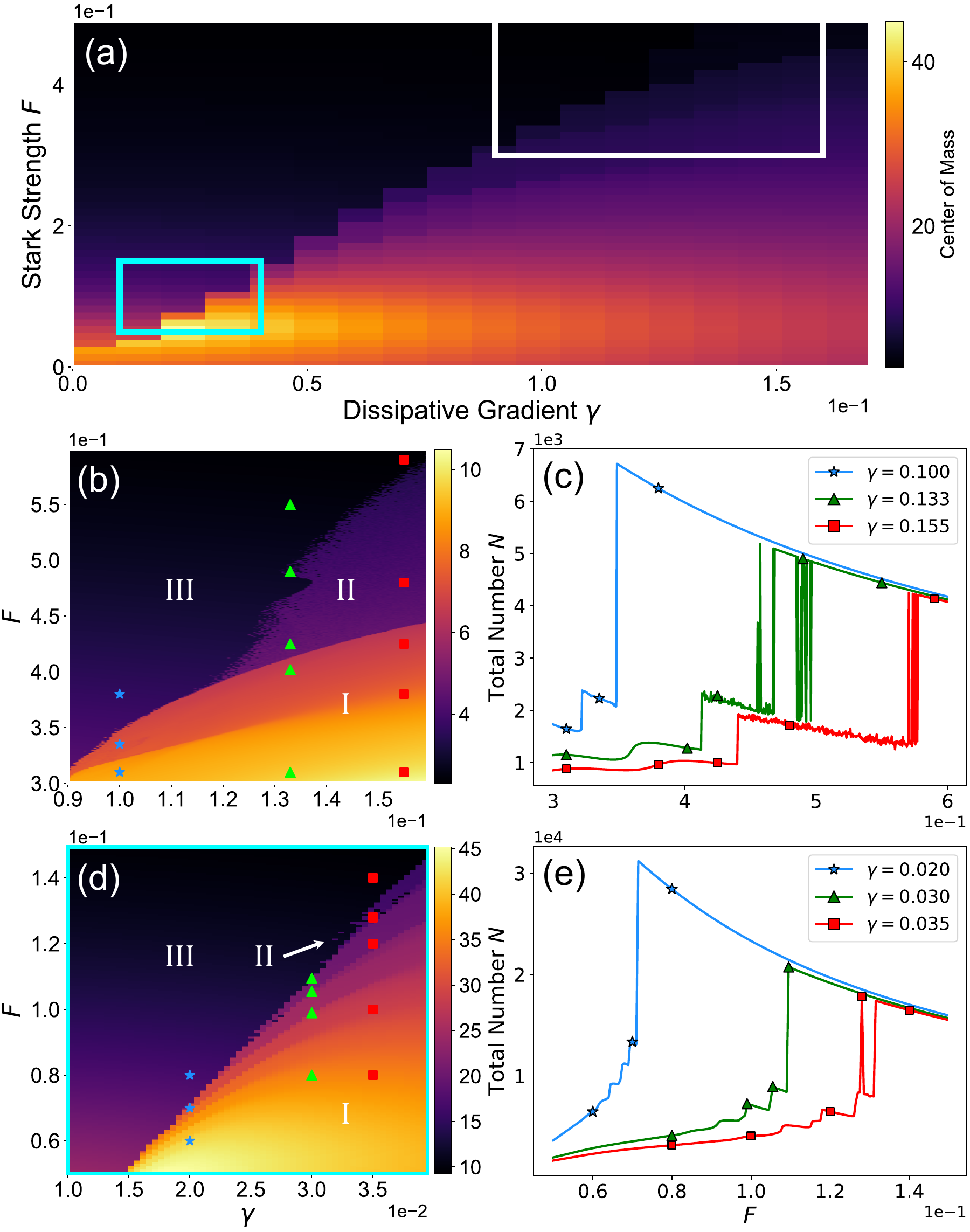}
  \caption{Dynamical phase diagram of the dissipation-selected front in the plane of detuning-ramp slope and dissipation-gradient strength, $(F, \gamma)$, characterized by the center of mass. (a) Global phase map revealing sharp transition boundaries. (b, d) High-resolution scans of the (b) high-$F$, high-$\gamma$ and (d) small-$F$, small-$\gamma$ sectors. These diagrams delineate three distinct phases: I (pinned), II (chaotic), and III (minimum-loss-localized). The pinned phase manifests a characteristic staircase structure, which is progressively smoothed by increasing dissipation gradient $\gamma$. (c, e) Total particle number $N$ versus $F$ at fixed $\gamma$ values corresponding to the sectors in (b) and (d). The traces illustrate the discrete steps, chaotic fluctuations, and abrupt jumps characteristic of the identified phases.}
  \label{fig:PatternTilt}
\end{figure}

\textit{Dynamical phases and depinning.}---We convert these representative trajectories into a dynamical phase diagram by scanning $(F,\gamma)$ and characterizing each long-time density distribution by its center of mass along the gradient direction,
\begin{equation}
    \mathrm{COM}_x = \frac{\sum_{x,y} x\,|\alpha_{x,y}|^2}{\sum_{x,y} |\alpha_{x,y}|^2}.
\end{equation}
The phase diagram in Figs.~\!\ref{fig:PatternTilt}(a,b,d) displays three domains separated by sharp transition boundaries. Phase I (pinned phase) supports a stable front localized at intermediate positions ($x_s>0$), appearing either uniform or pattern locked. Phase II (chaotic phase) sets in beyond the stability threshold, where the pinned front destabilizes and the system exhibits sustained spatiotemporal chaos, as illustrated in Fig.~\!\ref{fig:dynamics_time}(c). Phase III (minimum-loss-localized phase) is characterized by localization of the high-density region at the minimum-loss side ($x=0$), as depicted in Fig.~\!\ref{fig:dynamics_time}(d). The phase diagrams also show that transitions between domains follow more than one pathway. In some regions of parameter space, minimum-loss localization occurs before chaotic destabilization, giving a direct Phase I $\to$ Phase III transition; in others, pinned-front destabilization first produces spatiotemporal chaos and only later ends in the minimum-loss-localized phase, i.e., Phase I $\to$ Phase II $\to$ Phase III. The mechanism selecting these pathways is analyzed in~\cite{supplementary}.

While Phases II and III are structurally simple, the pinned phase contains a fine structure that exposes the lattice origin of the front. Its relocation is not continuous but proceeds via abrupt site-to-site jumps, manifesting as a stepwise response in the total particle number. This apparent quantization originates from the Peierls-Nabarro (PN) barrier~\cite{PhysRevA.69.053604,Jenkinson:2017aa}, which locks the front to lattice columns. We describe this pinning mechanism via the balance equation
\begin{equation}
\Delta\omega + F x_s - 4J = \varepsilon_{\mathrm{PN}}\sin(2\pi x_s) + \frac{4U\Omega^2}{\big(\kappa_0+\gamma x_s\big)^2},
\end{equation}
where the interplay between the detuning ramp and the periodic PN potential of amplitude $\varepsilon_{\mathrm{PN}}$ generates a ladder of depinning thresholds (see~\cite{supplementary} for details). As $F$ crosses successive thresholds, the pinned position $x_s$ advances by one lattice column. Increasing the dissipation gradient $\gamma$ modifies this staircase structure. The pinning plateaus widen because the resonant position becomes less sensitive to the ramp slope. Implicitly differentiating the pinning condition yields
\begin{equation}
\frac{\dd x_s}{\dd F}=-\frac{x_s}{F - R'(x_s)},\quad
R'(x)=-\frac{8U\Omega^2\gamma}{\big(\kappa_0+\gamma x\big)^3}<0,
\label{eq:dxdf_text}
\end{equation}
which confirms that a stronger gradient suppresses this response, thereby stabilizing the pinned state. The particle-number jump per step is governed by the upstream density,
\begin{equation}
\Delta N \approx L_y n_{\mathrm{up}}(x_s),
\label{eq:jump_text}
\end{equation}
where $L_y$ is the transverse system size and $n_{\mathrm{up}}(x)\propto \Omega^2/(\kappa_0+\gamma x)^2$ is the upstream density. A stronger dissipation gradient therefore erodes the jump height, and blurs the sharp steps into an effectively continuous response. The particle-number staircase is a macroscopic readout of column-by-column depinning of a nonlinear resonant front.

\begin{figure}[t]
\centering
\includegraphics[width=\linewidth]{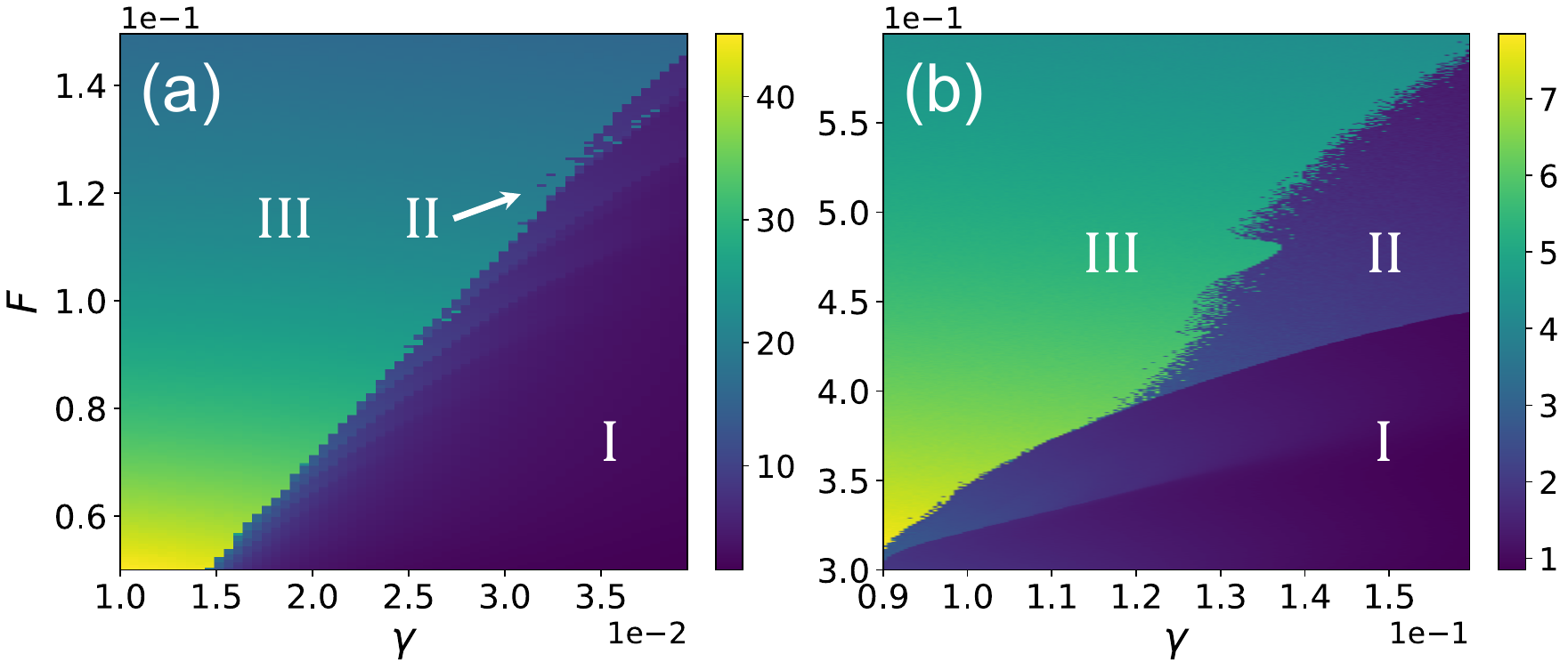}
\caption{Generalized imbalance $\mathcal{I}$ mapped in the plane of detuning-ramp slope and dissipation-gradient strength, $(F, \gamma)$. (a) and (b) correspond to the low- and high-$\gamma$ sectors shown in Figs.~\!\ref{fig:PatternTilt}(d, b), respectively. The imbalance landscape reproduces the intricate structure observed in the center-of-mass diagram, confirming the sharp boundaries between the pinned, chaotic, and minimum-loss-localized phases. Notably, the discrete terraces are clearly resolved, providing evidence that the staircase depinning and dynamical transitions are intrinsic features of the system.}
\label{fig:imbalance}
\end{figure}

Finally, we use a generalized imbalance, independent of the front coordinate, to test the phase boundaries. Following imbalance diagnostics of density memory~\cite{szdc-61nl,Mak:2024aa,Schreiber2015,PhysRevLett.119.260401}, we use the density-overlap diagnostic $\mathcal{I}(t)=\sum_j N_j(t)N_j(0)/\sum_j N_j^2(0)$, with $N_j(t)=|\alpha_j(t)|^2$; details in~\cite{supplementary}. The resulting phase map in Fig.~\!\ref{fig:imbalance} reproduces the $\mathrm{COM}_x$ boundaries and the characteristic terrace structure of the pinned phase, confirming that the regimes are not artifacts of a specific observable.

\textit{Conclusions.}---We have shown that patterned dissipation, when aligned with a Stark-induced detuning ramp, spatially selects nonlinear resonant fronts in a two-dimensional driven Bose-Hubbard lattice. The selected density front is pinned where the Kerr-shifted detuning vanishes, $\Delta_{\text{eff}}(x_s)=0$, and its Airy-like width is set by the balance between tunneling stiffness and the effective detuning slope. Sweeping the detuning-ramp slope and dissipation-gradient strength, we constructed a dynamical phase diagram of the resulting long-time states. The diagram contains a pinned phase where the front remains uniform or pattern locked, a chaotic phase driven by transverse instability, and a minimum-loss-localized phase where the high-density region is localized at the minimum-loss side. We further found that the pinned phase contains a quantized fine structure: lattice discreteness generates a Peierls-Nabarro depinning staircase in the total particle number, whose visibility is controlled by the dissipation gradient. Finally, the generalized imbalance provides a front-independent validation of the same phase boundaries. Overall, our work demonstrates that engineered dissipation structures offer a complementary dimension to Hamiltonian control, capable of dynamically regulating the stability and transport of nonlinear fronts. This strategy provides a route for interface engineering in programmable quantum simulators.

\textit{Acknowledgments}---We thank Dr. Yun-Hao Shi, and Prof. Dr. Luigi Amico for helpful discussions. This work was supported by the National Natural Science Foundation of China (Grants No. 92265207, No. T2121001, No. U25A6009, No. T2322030, No. 12122504, No. 12274142, and No. 12475017), QNMP (Grant No. 2021ZD0301800). 
% -------------------------------------------------------------

\providecommand{\noopsort}[1]{}\providecommand{\singleletter}[1]{#1}%


\begin{thebibliography}{71}%
\makeatletter
\providecommand \@ifxundefined [1]{%
 \@ifx{#1\undefined}
}%
\providecommand \@ifnum [1]{%
 \ifnum #1\expandafter \@firstoftwo
 \else \expandafter \@secondoftwo
 \fi
}%
\providecommand \@ifx [1]{%
 \ifx #1\expandafter \@firstoftwo
 \else \expandafter \@secondoftwo
 \fi
}%
\providecommand \natexlab [1]{#1}%
\providecommand \enquote  [1]{``#1''}%
\providecommand \bibnamefont  [1]{#1}%
\providecommand \bibfnamefont [1]{#1}%
\providecommand \citenamefont [1]{#1}%
\providecommand \href@noop [0]{\@secondoftwo}%
\providecommand \href [0]{\begingroup \@sanitize@url \@href}%
\providecommand \@href[1]{\@@startlink{#1}\@@href}%
\providecommand \@@href[1]{\endgroup#1\@@endlink}%
\providecommand \@sanitize@url [0]{\catcode `\\12\catcode `\$12\catcode `\&12\catcode `\#12\catcode `\^12\catcode `\_12\catcode `\%12\relax}%
\providecommand \@@startlink[1]{}%
\providecommand \@@endlink[0]{}%
\providecommand \url  [0]{\begingroup\@sanitize@url \@url }%
\providecommand \@url [1]{\endgroup\@href {#1}{\urlprefix }}%
\providecommand \urlprefix  [0]{URL }%
\providecommand \Eprint [0]{\href }%
\providecommand \doibase [0]{https://doi.org/}%
\providecommand \selectlanguage [0]{\@gobble}%
\providecommand \bibinfo  [0]{\@secondoftwo}%
\providecommand \bibfield  [0]{\@secondoftwo}%
\providecommand \translation [1]{[#1]}%
\providecommand \BibitemOpen [0]{}%
\providecommand \bibitemStop [0]{}%
\providecommand \bibitemNoStop [0]{.\EOS\space}%
\providecommand \EOS [0]{\spacefactor3000\relax}%
\providecommand \BibitemShut  [1]{\csname bibitem#1\endcsname}%
\let\auto@bib@innerbib\@empty
%</preamble>
\bibitem [{\citenamefont {Sieberer}\ \emph {et~al.}(2016)\citenamefont {Sieberer}, \citenamefont {Buchhold},\ and\ \citenamefont {Diehl}}]{Sieberer_2016}%
  \BibitemOpen
  \bibfield  {author} {\bibinfo {author} {\bibfnamefont {L.~M.}\ \bibnamefont {Sieberer}}, \bibinfo {author} {\bibfnamefont {M.}~\bibnamefont {Buchhold}},\ and\ \bibinfo {author} {\bibfnamefont {S.}~\bibnamefont {Diehl}},\ }\bibfield  {title} {\bibinfo {title} {Keldysh field theory for driven open quantum systems},\ }\href {https://doi.org/10.1088/0034-4885/79/9/096001} {\bibfield  {journal} {\bibinfo  {journal} {Reports on Progress in Physics}\ }\textbf {\bibinfo {volume} {79}},\ \bibinfo {pages} {096001} (\bibinfo {year} {2016})}\BibitemShut {NoStop}%
\bibitem [{\citenamefont {Noh}\ and\ \citenamefont {Angelakis}(2016)}]{Noh_2017}%
  \BibitemOpen
  \bibfield  {author} {\bibinfo {author} {\bibfnamefont {C.}~\bibnamefont {Noh}}\ and\ \bibinfo {author} {\bibfnamefont {D.~G.}\ \bibnamefont {Angelakis}},\ }\bibfield  {title} {\bibinfo {title} {Quantum simulations and many-body physics with light},\ }\href {https://doi.org/10.1088/0034-4885/80/1/016401} {\bibfield  {journal} {\bibinfo  {journal} {Reports on Progress in Physics}\ }\textbf {\bibinfo {volume} {80}},\ \bibinfo {pages} {016401} (\bibinfo {year} {2016})}\BibitemShut {NoStop}%
\bibitem [{\citenamefont {Dutta}\ \emph {et~al.}(2015)\citenamefont {Dutta}, \citenamefont {Gajda}, \citenamefont {Hauke}, \citenamefont {Lewenstein}, \citenamefont {L{\"u}hmann}, \citenamefont {Malomed}, \citenamefont {Sowi{\'n}ski},\ and\ \citenamefont {Zakrzewski}}]{Dutta_2015}%
  \BibitemOpen
  \bibfield  {author} {\bibinfo {author} {\bibfnamefont {O.}~\bibnamefont {Dutta}}, \bibinfo {author} {\bibfnamefont {M.}~\bibnamefont {Gajda}}, \bibinfo {author} {\bibfnamefont {P.}~\bibnamefont {Hauke}}, \bibinfo {author} {\bibfnamefont {M.}~\bibnamefont {Lewenstein}}, \bibinfo {author} {\bibfnamefont {D.-S.}\ \bibnamefont {L{\"u}hmann}}, \bibinfo {author} {\bibfnamefont {B.~A.}\ \bibnamefont {Malomed}}, \bibinfo {author} {\bibfnamefont {T.}~\bibnamefont {Sowi{\'n}ski}},\ and\ \bibinfo {author} {\bibfnamefont {J.}~\bibnamefont {Zakrzewski}},\ }\bibfield  {title} {\bibinfo {title} {Non-standard hubbard models in optical lattices: a review},\ }\href {https://doi.org/10.1088/0034-4885/78/6/066001} {\bibfield  {journal} {\bibinfo  {journal} {Reports on Progress in Physics}\ }\textbf {\bibinfo {volume} {78}},\ \bibinfo {pages} {066001} (\bibinfo {year} {2015})}\BibitemShut {NoStop}%
\bibitem [{\citenamefont {Fitzpatrick}\ \emph {et~al.}(2017)\citenamefont {Fitzpatrick}, \citenamefont {Sundaresan}, \citenamefont {Li}, \citenamefont {Koch},\ and\ \citenamefont {Houck}}]{PhysRevX.7.011016}%
  \BibitemOpen
  \bibfield  {author} {\bibinfo {author} {\bibfnamefont {M.}~\bibnamefont {Fitzpatrick}}, \bibinfo {author} {\bibfnamefont {N.~M.}\ \bibnamefont {Sundaresan}}, \bibinfo {author} {\bibfnamefont {A.~C.~Y.}\ \bibnamefont {Li}}, \bibinfo {author} {\bibfnamefont {J.}~\bibnamefont {Koch}},\ and\ \bibinfo {author} {\bibfnamefont {A.~A.}\ \bibnamefont {Houck}},\ }\bibfield  {title} {\bibinfo {title} {Observation of a dissipative phase transition in a one-dimensional circuit qed lattice},\ }\href {https://doi.org/10.1103/PhysRevX.7.011016} {\bibfield  {journal} {\bibinfo  {journal} {Phys. Rev. X}\ }\textbf {\bibinfo {volume} {7}},\ \bibinfo {pages} {011016} (\bibinfo {year} {2017})}\BibitemShut {NoStop}%
\bibitem [{\citenamefont {Fink}\ \emph {et~al.}(2017)\citenamefont {Fink}, \citenamefont {Dombi}, \citenamefont {Vukics}, \citenamefont {Wallraff},\ and\ \citenamefont {Domokos}}]{PhysRevX.7.011012}%
  \BibitemOpen
  \bibfield  {author} {\bibinfo {author} {\bibfnamefont {J.~M.}\ \bibnamefont {Fink}}, \bibinfo {author} {\bibfnamefont {A.}~\bibnamefont {Dombi}}, \bibinfo {author} {\bibfnamefont {A.}~\bibnamefont {Vukics}}, \bibinfo {author} {\bibfnamefont {A.}~\bibnamefont {Wallraff}},\ and\ \bibinfo {author} {\bibfnamefont {P.}~\bibnamefont {Domokos}},\ }\bibfield  {title} {\bibinfo {title} {Observation of the photon-blockade breakdown phase transition},\ }\href {https://doi.org/10.1103/PhysRevX.7.011012} {\bibfield  {journal} {\bibinfo  {journal} {Phys. Rev. X}\ }\textbf {\bibinfo {volume} {7}},\ \bibinfo {pages} {011012} (\bibinfo {year} {2017})}\BibitemShut {NoStop}%
\bibitem [{\citenamefont {Minganti}\ \emph {et~al.}(2018)\citenamefont {Minganti}, \citenamefont {Biella}, \citenamefont {Bartolo},\ and\ \citenamefont {Ciuti}}]{PhysRevA.98.042118}%
  \BibitemOpen
  \bibfield  {author} {\bibinfo {author} {\bibfnamefont {F.}~\bibnamefont {Minganti}}, \bibinfo {author} {\bibfnamefont {A.}~\bibnamefont {Biella}}, \bibinfo {author} {\bibfnamefont {N.}~\bibnamefont {Bartolo}},\ and\ \bibinfo {author} {\bibfnamefont {C.}~\bibnamefont {Ciuti}},\ }\bibfield  {title} {\bibinfo {title} {Spectral theory of liouvillians for dissipative phase transitions},\ }\href {https://doi.org/10.1103/PhysRevA.98.042118} {\bibfield  {journal} {\bibinfo  {journal} {Phys. Rev. A}\ }\textbf {\bibinfo {volume} {98}},\ \bibinfo {pages} {042118} (\bibinfo {year} {2018})}\BibitemShut {NoStop}%
\bibitem [{\citenamefont {Rodriguez}\ \emph {et~al.}(2017)\citenamefont {Rodriguez}, \citenamefont {Casteels}, \citenamefont {Storme}, \citenamefont {Carlon~Zambon}, \citenamefont {Sagnes}, \citenamefont {Le~Gratiet}, \citenamefont {Galopin}, \citenamefont {Lema\^{\i}tre}, \citenamefont {Amo}, \citenamefont {Ciuti},\ and\ \citenamefont {Bloch}}]{PhysRevLett.118.247402}%
  \BibitemOpen
  \bibfield  {author} {\bibinfo {author} {\bibfnamefont {S.~R.~K.}\ \bibnamefont {Rodriguez}}, \bibinfo {author} {\bibfnamefont {W.}~\bibnamefont {Casteels}}, \bibinfo {author} {\bibfnamefont {F.}~\bibnamefont {Storme}}, \bibinfo {author} {\bibfnamefont {N.}~\bibnamefont {Carlon~Zambon}}, \bibinfo {author} {\bibfnamefont {I.}~\bibnamefont {Sagnes}}, \bibinfo {author} {\bibfnamefont {L.}~\bibnamefont {Le~Gratiet}}, \bibinfo {author} {\bibfnamefont {E.}~\bibnamefont {Galopin}}, \bibinfo {author} {\bibfnamefont {A.}~\bibnamefont {Lema\^{\i}tre}}, \bibinfo {author} {\bibfnamefont {A.}~\bibnamefont {Amo}}, \bibinfo {author} {\bibfnamefont {C.}~\bibnamefont {Ciuti}},\ and\ \bibinfo {author} {\bibfnamefont {J.}~\bibnamefont {Bloch}},\ }\bibfield  {title} {\bibinfo {title} {Probing a dissipative phase transition via dynamical optical hysteresis},\ }\href {https://doi.org/10.1103/PhysRevLett.118.247402} {\bibfield  {journal} {\bibinfo  {journal} {Phys. Rev. Lett.}\ }\textbf {\bibinfo {volume} {118}},\ \bibinfo {pages}
  {247402} (\bibinfo {year} {2017})}\BibitemShut {NoStop}%
\bibitem [{\citenamefont {Letscher}\ \emph {et~al.}(2017)\citenamefont {Letscher}, \citenamefont {Thomas}, \citenamefont {Niederpr\"um}, \citenamefont {Fleischhauer},\ and\ \citenamefont {Ott}}]{PhysRevX.7.021020}%
  \BibitemOpen
  \bibfield  {author} {\bibinfo {author} {\bibfnamefont {F.}~\bibnamefont {Letscher}}, \bibinfo {author} {\bibfnamefont {O.}~\bibnamefont {Thomas}}, \bibinfo {author} {\bibfnamefont {T.}~\bibnamefont {Niederpr\"um}}, \bibinfo {author} {\bibfnamefont {M.}~\bibnamefont {Fleischhauer}},\ and\ \bibinfo {author} {\bibfnamefont {H.}~\bibnamefont {Ott}},\ }\bibfield  {title} {\bibinfo {title} {Bistability versus metastability in driven dissipative rydberg gases},\ }\href {https://doi.org/10.1103/PhysRevX.7.021020} {\bibfield  {journal} {\bibinfo  {journal} {Phys. Rev. X}\ }\textbf {\bibinfo {volume} {7}},\ \bibinfo {pages} {021020} (\bibinfo {year} {2017})}\BibitemShut {NoStop}%
\bibitem [{\citenamefont {Harari}\ \emph {et~al.}(2018)\citenamefont {Harari}, \citenamefont {Bandres}, \citenamefont {Lumer}, \citenamefont {Rechtsman}, \citenamefont {Chong}, \citenamefont {Khajavikhan}, \citenamefont {Christodoulides},\ and\ \citenamefont {Segev}}]{aar4003}%
  \BibitemOpen
  \bibfield  {author} {\bibinfo {author} {\bibfnamefont {G.}~\bibnamefont {Harari}}, \bibinfo {author} {\bibfnamefont {M.~A.}\ \bibnamefont {Bandres}}, \bibinfo {author} {\bibfnamefont {Y.}~\bibnamefont {Lumer}}, \bibinfo {author} {\bibfnamefont {M.~C.}\ \bibnamefont {Rechtsman}}, \bibinfo {author} {\bibfnamefont {Y.~D.}\ \bibnamefont {Chong}}, \bibinfo {author} {\bibfnamefont {M.}~\bibnamefont {Khajavikhan}}, \bibinfo {author} {\bibfnamefont {D.~N.}\ \bibnamefont {Christodoulides}},\ and\ \bibinfo {author} {\bibfnamefont {M.}~\bibnamefont {Segev}},\ }\bibfield  {title} {\bibinfo {title} {Topological insulator laser: Theory},\ }\href {https://doi.org/10.1126/science.aar4003} {\bibfield  {journal} {\bibinfo  {journal} {Science}\ }\textbf {\bibinfo {volume} {359}},\ \bibinfo {pages} {eaar4003} (\bibinfo {year} {2018})},\ \Eprint {https://arxiv.org/abs/https://www.science.org/doi/pdf/10.1126/science.aar4003} {https://www.science.org/doi/pdf/10.1126/science.aar4003} \BibitemShut {NoStop}%
\bibitem [{\citenamefont {Owen}\ \emph {et~al.}(2018)\citenamefont {Owen}, \citenamefont {Jin}, \citenamefont {Rossini}, \citenamefont {Fazio},\ and\ \citenamefont {Hartmann}}]{Owen_2018}%
  \BibitemOpen
  \bibfield  {author} {\bibinfo {author} {\bibfnamefont {E.~T.}\ \bibnamefont {Owen}}, \bibinfo {author} {\bibfnamefont {J.}~\bibnamefont {Jin}}, \bibinfo {author} {\bibfnamefont {D.}~\bibnamefont {Rossini}}, \bibinfo {author} {\bibfnamefont {R.}~\bibnamefont {Fazio}},\ and\ \bibinfo {author} {\bibfnamefont {M.~J.}\ \bibnamefont {Hartmann}},\ }\bibfield  {title} {\bibinfo {title} {Quantum correlations and limit cycles in the driven-dissipative heisenberg lattice},\ }\href {https://doi.org/10.1088/1367-2630/aab7d3} {\bibfield  {journal} {\bibinfo  {journal} {New Journal of Physics}\ }\textbf {\bibinfo {volume} {20}},\ \bibinfo {pages} {045004} (\bibinfo {year} {2018})}\BibitemShut {NoStop}%
\bibitem [{\citenamefont {Rota}\ \emph {et~al.}(2017)\citenamefont {Rota}, \citenamefont {Storme}, \citenamefont {Bartolo}, \citenamefont {Fazio},\ and\ \citenamefont {Ciuti}}]{PhysRevB.95.134431}%
  \BibitemOpen
  \bibfield  {author} {\bibinfo {author} {\bibfnamefont {R.}~\bibnamefont {Rota}}, \bibinfo {author} {\bibfnamefont {F.}~\bibnamefont {Storme}}, \bibinfo {author} {\bibfnamefont {N.}~\bibnamefont {Bartolo}}, \bibinfo {author} {\bibfnamefont {R.}~\bibnamefont {Fazio}},\ and\ \bibinfo {author} {\bibfnamefont {C.}~\bibnamefont {Ciuti}},\ }\bibfield  {title} {\bibinfo {title} {Critical behavior of dissipative two-dimensional spin lattices},\ }\href {https://doi.org/10.1103/PhysRevB.95.134431} {\bibfield  {journal} {\bibinfo  {journal} {Phys. Rev. B}\ }\textbf {\bibinfo {volume} {95}},\ \bibinfo {pages} {134431} (\bibinfo {year} {2017})}\BibitemShut {NoStop}%
\bibitem [{\citenamefont {Gersch}\ and\ \citenamefont {Knollman}(1963)}]{PhysRev.129.959}%
  \BibitemOpen
  \bibfield  {author} {\bibinfo {author} {\bibfnamefont {H.~A.}\ \bibnamefont {Gersch}}\ and\ \bibinfo {author} {\bibfnamefont {G.~C.}\ \bibnamefont {Knollman}},\ }\bibfield  {title} {\bibinfo {title} {Quantum cell model for bosons},\ }\href {https://doi.org/10.1103/PhysRev.129.959} {\bibfield  {journal} {\bibinfo  {journal} {Phys. Rev.}\ }\textbf {\bibinfo {volume} {129}},\ \bibinfo {pages} {959} (\bibinfo {year} {1963})}\BibitemShut {NoStop}%
\bibitem [{\citenamefont {Fisher}\ \emph {et~al.}(1989)\citenamefont {Fisher}, \citenamefont {Weichman}, \citenamefont {Grinstein},\ and\ \citenamefont {Fisher}}]{PhysRevB.40.546}%
  \BibitemOpen
  \bibfield  {author} {\bibinfo {author} {\bibfnamefont {M.~P.~A.}\ \bibnamefont {Fisher}}, \bibinfo {author} {\bibfnamefont {P.~B.}\ \bibnamefont {Weichman}}, \bibinfo {author} {\bibfnamefont {G.}~\bibnamefont {Grinstein}},\ and\ \bibinfo {author} {\bibfnamefont {D.~S.}\ \bibnamefont {Fisher}},\ }\bibfield  {title} {\bibinfo {title} {Boson localization and the superfluid-insulator transition},\ }\href {https://doi.org/10.1103/PhysRevB.40.546} {\bibfield  {journal} {\bibinfo  {journal} {Phys. Rev. B}\ }\textbf {\bibinfo {volume} {40}},\ \bibinfo {pages} {546} (\bibinfo {year} {1989})}\BibitemShut {NoStop}%
\bibitem [{\citenamefont {Kordas}\ \emph {et~al.}(2015)\citenamefont {Kordas}, \citenamefont {Witthaut}, \citenamefont {Buonsante}, \citenamefont {Vezzani}, \citenamefont {Burioni}, \citenamefont {Karanikas},\ and\ \citenamefont {Wimberger}}]{Kordas:2015aa}%
  \BibitemOpen
  \bibfield  {author} {\bibinfo {author} {\bibfnamefont {G.}~\bibnamefont {Kordas}}, \bibinfo {author} {\bibfnamefont {D.}~\bibnamefont {Witthaut}}, \bibinfo {author} {\bibfnamefont {P.}~\bibnamefont {Buonsante}}, \bibinfo {author} {\bibfnamefont {A.}~\bibnamefont {Vezzani}}, \bibinfo {author} {\bibfnamefont {R.}~\bibnamefont {Burioni}}, \bibinfo {author} {\bibfnamefont {A.~I.}\ \bibnamefont {Karanikas}},\ and\ \bibinfo {author} {\bibfnamefont {S.}~\bibnamefont {Wimberger}},\ }\bibfield  {title} {\bibinfo {title} {The dissipative bose-hubbard model},\ }\href {https://doi.org/10.1140/epjst/e2015-02528-2} {\bibfield  {journal} {\bibinfo  {journal} {The European Physical Journal Special Topics}\ }\textbf {\bibinfo {volume} {224}},\ \bibinfo {pages} {2127} (\bibinfo {year} {2015})}\BibitemShut {NoStop}%
\bibitem [{\citenamefont {Ekman}\ and\ \citenamefont {Bergholtz}(2024)}]{PhysRevResearch.6.L032067}%
  \BibitemOpen
  \bibfield  {author} {\bibinfo {author} {\bibfnamefont {C.}~\bibnamefont {Ekman}}\ and\ \bibinfo {author} {\bibfnamefont {E.~J.}\ \bibnamefont {Bergholtz}},\ }\bibfield  {title} {\bibinfo {title} {Liouvillian skin effects and fragmented condensates in an integrable dissipative bose-hubbard model},\ }\href {https://doi.org/10.1103/PhysRevResearch.6.L032067} {\bibfield  {journal} {\bibinfo  {journal} {Phys. Rev. Res.}\ }\textbf {\bibinfo {volume} {6}},\ \bibinfo {pages} {L032067} (\bibinfo {year} {2024})}\BibitemShut {NoStop}%
\bibitem [{\citenamefont {Tomadin}\ \emph {et~al.}(2010)\citenamefont {Tomadin}, \citenamefont {Giovannetti}, \citenamefont {Fazio}, \citenamefont {Gerace}, \citenamefont {Carusotto}, \citenamefont {T\"ureci},\ and\ \citenamefont {Imamoglu}}]{PhysRevA.81.061801}%
  \BibitemOpen
  \bibfield  {author} {\bibinfo {author} {\bibfnamefont {A.}~\bibnamefont {Tomadin}}, \bibinfo {author} {\bibfnamefont {V.}~\bibnamefont {Giovannetti}}, \bibinfo {author} {\bibfnamefont {R.}~\bibnamefont {Fazio}}, \bibinfo {author} {\bibfnamefont {D.}~\bibnamefont {Gerace}}, \bibinfo {author} {\bibfnamefont {I.}~\bibnamefont {Carusotto}}, \bibinfo {author} {\bibfnamefont {H.~E.}\ \bibnamefont {T\"ureci}},\ and\ \bibinfo {author} {\bibfnamefont {A.}~\bibnamefont {Imamoglu}},\ }\bibfield  {title} {\bibinfo {title} {Signatures of the superfluid-insulator phase transition in laser-driven dissipative nonlinear cavity arrays},\ }\href {https://doi.org/10.1103/PhysRevA.81.061801} {\bibfield  {journal} {\bibinfo  {journal} {Phys. Rev. A}\ }\textbf {\bibinfo {volume} {81}},\ \bibinfo {pages} {061801} (\bibinfo {year} {2010})}\BibitemShut {NoStop}%
\bibitem [{\citenamefont {Jouanny}\ \emph {et~al.}(2025)\citenamefont {Jouanny}, \citenamefont {Frasca}, \citenamefont {Weibel}, \citenamefont {Peyruchat}, \citenamefont {Scigliuzzo}, \citenamefont {Oppliger}, \citenamefont {De~Palma}, \citenamefont {Sbroggi{\`o}}, \citenamefont {Beaulieu}, \citenamefont {Zilberberg},\ and\ \citenamefont {Scarlino}}]{Jouanny:2025aa}%
  \BibitemOpen
  \bibfield  {author} {\bibinfo {author} {\bibfnamefont {V.}~\bibnamefont {Jouanny}}, \bibinfo {author} {\bibfnamefont {S.}~\bibnamefont {Frasca}}, \bibinfo {author} {\bibfnamefont {V.~J.}\ \bibnamefont {Weibel}}, \bibinfo {author} {\bibfnamefont {l.}~\bibnamefont {Peyruchat}}, \bibinfo {author} {\bibfnamefont {M.}~\bibnamefont {Scigliuzzo}}, \bibinfo {author} {\bibfnamefont {F.}~\bibnamefont {Oppliger}}, \bibinfo {author} {\bibfnamefont {F.}~\bibnamefont {De~Palma}}, \bibinfo {author} {\bibfnamefont {D.}~\bibnamefont {Sbroggi{\`o}}}, \bibinfo {author} {\bibfnamefont {G.}~\bibnamefont {Beaulieu}}, \bibinfo {author} {\bibfnamefont {O.}~\bibnamefont {Zilberberg}},\ and\ \bibinfo {author} {\bibfnamefont {P.}~\bibnamefont {Scarlino}},\ }\bibfield  {title} {\bibinfo {title} {High kinetic inductance cavity arrays for compact band engineering and topology-based disorder meters},\ }\href {https://doi.org/10.1038/s41467-025-58595-8} {\bibfield  {journal} {\bibinfo  {journal} {Nature Communications}\ }\textbf {\bibinfo
  {volume} {16}},\ \bibinfo {pages} {3396} (\bibinfo {year} {2025})}\BibitemShut {NoStop}%
\bibitem [{\citenamefont {Wu}\ \emph {et~al.}(2011)\citenamefont {Wu}, \citenamefont {Gao}, \citenamefont {Deng}, \citenamefont {Dai}, \citenamefont {Chen},\ and\ \citenamefont {Li}}]{PhysRevA.84.043827}%
  \BibitemOpen
  \bibfield  {author} {\bibinfo {author} {\bibfnamefont {C.-W.}\ \bibnamefont {Wu}}, \bibinfo {author} {\bibfnamefont {M.}~\bibnamefont {Gao}}, \bibinfo {author} {\bibfnamefont {Z.-J.}\ \bibnamefont {Deng}}, \bibinfo {author} {\bibfnamefont {H.-Y.}\ \bibnamefont {Dai}}, \bibinfo {author} {\bibfnamefont {P.-X.}\ \bibnamefont {Chen}},\ and\ \bibinfo {author} {\bibfnamefont {C.-Z.}\ \bibnamefont {Li}},\ }\bibfield  {title} {\bibinfo {title} {Quantum phase transition of light in a one-dimensional photon-hopping-controllable resonator array},\ }\href {https://doi.org/10.1103/PhysRevA.84.043827} {\bibfield  {journal} {\bibinfo  {journal} {Phys. Rev. A}\ }\textbf {\bibinfo {volume} {84}},\ \bibinfo {pages} {043827} (\bibinfo {year} {2011})}\BibitemShut {NoStop}%
\bibitem [{\citenamefont {Fedorov}\ \emph {et~al.}(2021)\citenamefont {Fedorov}, \citenamefont {Remizov}, \citenamefont {Shapiro}, \citenamefont {Pogosov}, \citenamefont {Egorova}, \citenamefont {Tsitsilin}, \citenamefont {Andronik}, \citenamefont {Dobronosova}, \citenamefont {Rodionov}, \citenamefont {Astafiev},\ and\ \citenamefont {Ustinov}}]{PhysRevLett.126.180503}%
  \BibitemOpen
  \bibfield  {author} {\bibinfo {author} {\bibfnamefont {G.~P.}\ \bibnamefont {Fedorov}}, \bibinfo {author} {\bibfnamefont {S.~V.}\ \bibnamefont {Remizov}}, \bibinfo {author} {\bibfnamefont {D.~S.}\ \bibnamefont {Shapiro}}, \bibinfo {author} {\bibfnamefont {W.~V.}\ \bibnamefont {Pogosov}}, \bibinfo {author} {\bibfnamefont {E.}~\bibnamefont {Egorova}}, \bibinfo {author} {\bibfnamefont {I.}~\bibnamefont {Tsitsilin}}, \bibinfo {author} {\bibfnamefont {M.}~\bibnamefont {Andronik}}, \bibinfo {author} {\bibfnamefont {A.~A.}\ \bibnamefont {Dobronosova}}, \bibinfo {author} {\bibfnamefont {I.~A.}\ \bibnamefont {Rodionov}}, \bibinfo {author} {\bibfnamefont {O.~V.}\ \bibnamefont {Astafiev}},\ and\ \bibinfo {author} {\bibfnamefont {A.~V.}\ \bibnamefont {Ustinov}},\ }\bibfield  {title} {\bibinfo {title} {Photon transport in a bose-hubbard chain of superconducting artificial atoms},\ }\href {https://doi.org/10.1103/PhysRevLett.126.180503} {\bibfield  {journal} {\bibinfo  {journal} {Phys. Rev. Lett.}\ }\textbf {\bibinfo
  {volume} {126}},\ \bibinfo {pages} {180503} (\bibinfo {year} {2021})}\BibitemShut {NoStop}%
\bibitem [{\citenamefont {Zhao}\ \emph {et~al.}(2020)\citenamefont {Zhao}, \citenamefont {Vovrosh}, \citenamefont {Mintert},\ and\ \citenamefont {Knolle}}]{PhysRevLett.124.160604}%
  \BibitemOpen
  \bibfield  {author} {\bibinfo {author} {\bibfnamefont {H.}~\bibnamefont {Zhao}}, \bibinfo {author} {\bibfnamefont {J.}~\bibnamefont {Vovrosh}}, \bibinfo {author} {\bibfnamefont {F.}~\bibnamefont {Mintert}},\ and\ \bibinfo {author} {\bibfnamefont {J.}~\bibnamefont {Knolle}},\ }\bibfield  {title} {\bibinfo {title} {Quantum many-body scars in optical lattices},\ }\href {https://doi.org/10.1103/PhysRevLett.124.160604} {\bibfield  {journal} {\bibinfo  {journal} {Phys. Rev. Lett.}\ }\textbf {\bibinfo {volume} {124}},\ \bibinfo {pages} {160604} (\bibinfo {year} {2020})}\BibitemShut {NoStop}%
\bibitem [{\citenamefont {Rossini}\ \emph {et~al.}(2013)\citenamefont {Rossini}, \citenamefont {Gibertini}, \citenamefont {Giovannetti},\ and\ \citenamefont {Fazio}}]{PhysRevB.87.085131}%
  \BibitemOpen
  \bibfield  {author} {\bibinfo {author} {\bibfnamefont {D.}~\bibnamefont {Rossini}}, \bibinfo {author} {\bibfnamefont {M.}~\bibnamefont {Gibertini}}, \bibinfo {author} {\bibfnamefont {V.}~\bibnamefont {Giovannetti}},\ and\ \bibinfo {author} {\bibfnamefont {R.}~\bibnamefont {Fazio}},\ }\bibfield  {title} {\bibinfo {title} {Topological pumping in the one-dimensional bose-hubbard model},\ }\href {https://doi.org/10.1103/PhysRevB.87.085131} {\bibfield  {journal} {\bibinfo  {journal} {Phys. Rev. B}\ }\textbf {\bibinfo {volume} {87}},\ \bibinfo {pages} {085131} (\bibinfo {year} {2013})}\BibitemShut {NoStop}%
\bibitem [{\citenamefont {Vicentini}\ \emph {et~al.}(2018)\citenamefont {Vicentini}, \citenamefont {Minganti}, \citenamefont {Rota}, \citenamefont {Orso},\ and\ \citenamefont {Ciuti}}]{PhysRevA.97.013853}%
  \BibitemOpen
  \bibfield  {author} {\bibinfo {author} {\bibfnamefont {F.}~\bibnamefont {Vicentini}}, \bibinfo {author} {\bibfnamefont {F.}~\bibnamefont {Minganti}}, \bibinfo {author} {\bibfnamefont {R.}~\bibnamefont {Rota}}, \bibinfo {author} {\bibfnamefont {G.}~\bibnamefont {Orso}},\ and\ \bibinfo {author} {\bibfnamefont {C.}~\bibnamefont {Ciuti}},\ }\bibfield  {title} {\bibinfo {title} {Critical slowing down in driven-dissipative bose-hubbard lattices},\ }\href {https://doi.org/10.1103/PhysRevA.97.013853} {\bibfield  {journal} {\bibinfo  {journal} {Phys. Rev. A}\ }\textbf {\bibinfo {volume} {97}},\ \bibinfo {pages} {013853} (\bibinfo {year} {2018})}\BibitemShut {NoStop}%
\bibitem [{\citenamefont {Roberts}\ and\ \citenamefont {Clerk}(2023)}]{PhysRevLett.130.063601}%
  \BibitemOpen
  \bibfield  {author} {\bibinfo {author} {\bibfnamefont {D.}~\bibnamefont {Roberts}}\ and\ \bibinfo {author} {\bibfnamefont {A.~A.}\ \bibnamefont {Clerk}},\ }\bibfield  {title} {\bibinfo {title} {Competition between two-photon driving, dissipation, and interactions in bosonic lattice models: An exact solution},\ }\href {https://doi.org/10.1103/PhysRevLett.130.063601} {\bibfield  {journal} {\bibinfo  {journal} {Phys. Rev. Lett.}\ }\textbf {\bibinfo {volume} {130}},\ \bibinfo {pages} {063601} (\bibinfo {year} {2023})}\BibitemShut {NoStop}%
\bibitem [{\citenamefont {Miyazaki}\ \emph {et~al.}(2003)\citenamefont {Miyazaki}, \citenamefont {Takahide}, \citenamefont {Kanda},\ and\ \citenamefont {Ootuka}}]{MIYAZAKI200341}%
  \BibitemOpen
  \bibfield  {author} {\bibinfo {author} {\bibfnamefont {H.}~\bibnamefont {Miyazaki}}, \bibinfo {author} {\bibfnamefont {Y.}~\bibnamefont {Takahide}}, \bibinfo {author} {\bibfnamefont {A.}~\bibnamefont {Kanda}},\ and\ \bibinfo {author} {\bibfnamefont {Y.}~\bibnamefont {Ootuka}},\ }\bibfield  {title} {\bibinfo {title} {Quantum fluctuations and dissipative phase transition in one-dimensional josephson junction arrays},\ }\href {https://doi.org/https://doi.org/10.1016/S1386-9477(02)00949-9} {\bibfield  {journal} {\bibinfo  {journal} {Physica E: Low-dimensional Systems and Nanostructures}\ }\textbf {\bibinfo {volume} {18}},\ \bibinfo {pages} {41} (\bibinfo {year} {2003})}\BibitemShut {NoStop}%
\bibitem [{\citenamefont {Ceulemans}\ and\ \citenamefont {Wouters}(2023)}]{ceulemans2023nonequilibrium}%
  \BibitemOpen
  \bibfield  {author} {\bibinfo {author} {\bibfnamefont {R.}~\bibnamefont {Ceulemans}}\ and\ \bibinfo {author} {\bibfnamefont {M.}~\bibnamefont {Wouters}},\ }\bibfield  {title} {\bibinfo {title} {Nonequilibrium steady states and critical slowing down in the dissipative bose-hubbard model},\ }\href {https://doi.org/10.1103/PhysRevA.108.013314} {\bibfield  {journal} {\bibinfo  {journal} {Phys. Rev. A}\ }\textbf {\bibinfo {volume} {108}},\ \bibinfo {pages} {013314} (\bibinfo {year} {2023})}\BibitemShut {NoStop}%
\bibitem [{\citenamefont {Casteels}\ \emph {et~al.}(2017)\citenamefont {Casteels}, \citenamefont {Fazio},\ and\ \citenamefont {Ciuti}}]{PhysRevA.95.012128}%
  \BibitemOpen
  \bibfield  {author} {\bibinfo {author} {\bibfnamefont {W.}~\bibnamefont {Casteels}}, \bibinfo {author} {\bibfnamefont {R.}~\bibnamefont {Fazio}},\ and\ \bibinfo {author} {\bibfnamefont {C.}~\bibnamefont {Ciuti}},\ }\bibfield  {title} {\bibinfo {title} {Critical dynamical properties of a first-order dissipative phase transition},\ }\href {https://doi.org/10.1103/PhysRevA.95.012128} {\bibfield  {journal} {\bibinfo  {journal} {Phys. Rev. A}\ }\textbf {\bibinfo {volume} {95}},\ \bibinfo {pages} {012128} (\bibinfo {year} {2017})}\BibitemShut {NoStop}%
\bibitem [{\citenamefont {Biella}\ \emph {et~al.}(2017)\citenamefont {Biella}, \citenamefont {Storme}, \citenamefont {Lebreuilly}, \citenamefont {Rossini}, \citenamefont {Fazio}, \citenamefont {Carusotto},\ and\ \citenamefont {Ciuti}}]{PhysRevA.96.023839}%
  \BibitemOpen
  \bibfield  {author} {\bibinfo {author} {\bibfnamefont {A.}~\bibnamefont {Biella}}, \bibinfo {author} {\bibfnamefont {F.}~\bibnamefont {Storme}}, \bibinfo {author} {\bibfnamefont {J.}~\bibnamefont {Lebreuilly}}, \bibinfo {author} {\bibfnamefont {D.}~\bibnamefont {Rossini}}, \bibinfo {author} {\bibfnamefont {R.}~\bibnamefont {Fazio}}, \bibinfo {author} {\bibfnamefont {I.}~\bibnamefont {Carusotto}},\ and\ \bibinfo {author} {\bibfnamefont {C.}~\bibnamefont {Ciuti}},\ }\bibfield  {title} {\bibinfo {title} {Phase diagram of incoherently driven strongly correlated photonic lattices},\ }\href {https://doi.org/10.1103/PhysRevA.96.023839} {\bibfield  {journal} {\bibinfo  {journal} {Phys. Rev. A}\ }\textbf {\bibinfo {volume} {96}},\ \bibinfo {pages} {023839} (\bibinfo {year} {2017})}\BibitemShut {NoStop}%
\bibitem [{\citenamefont {Le~Boit\'e}\ \emph {et~al.}(2014)\citenamefont {Le~Boit\'e}, \citenamefont {Orso},\ and\ \citenamefont {Ciuti}}]{PhysRevA.90.063821}%
  \BibitemOpen
  \bibfield  {author} {\bibinfo {author} {\bibfnamefont {A.}~\bibnamefont {Le~Boit\'e}}, \bibinfo {author} {\bibfnamefont {G.}~\bibnamefont {Orso}},\ and\ \bibinfo {author} {\bibfnamefont {C.}~\bibnamefont {Ciuti}},\ }\bibfield  {title} {\bibinfo {title} {Bose-hubbard model: Relation between driven-dissipative steady states and equilibrium quantum phases},\ }\href {https://doi.org/10.1103/PhysRevA.90.063821} {\bibfield  {journal} {\bibinfo  {journal} {Phys. Rev. A}\ }\textbf {\bibinfo {volume} {90}},\ \bibinfo {pages} {063821} (\bibinfo {year} {2014})}\BibitemShut {NoStop}%
\bibitem [{\citenamefont {Casteels}\ and\ \citenamefont {Wouters}(2017)}]{PhysRevA.95.043833}%
  \BibitemOpen
  \bibfield  {author} {\bibinfo {author} {\bibfnamefont {W.}~\bibnamefont {Casteels}}\ and\ \bibinfo {author} {\bibfnamefont {M.}~\bibnamefont {Wouters}},\ }\bibfield  {title} {\bibinfo {title} {Optically bistable driven-dissipative bose-hubbard dimer: Gutzwiller approaches and entanglement},\ }\href {https://doi.org/10.1103/PhysRevA.95.043833} {\bibfield  {journal} {\bibinfo  {journal} {Phys. Rev. A}\ }\textbf {\bibinfo {volume} {95}},\ \bibinfo {pages} {043833} (\bibinfo {year} {2017})}\BibitemShut {NoStop}%
\bibitem [{\citenamefont {Rodriguez}\ \emph {et~al.}(2016)\citenamefont {Rodriguez}, \citenamefont {Amo}, \citenamefont {Sagnes}, \citenamefont {Le~Gratiet}, \citenamefont {Galopin}, \citenamefont {Lema{\^\i}tre},\ and\ \citenamefont {Bloch}}]{Rodriguez:2016aa}%
  \BibitemOpen
  \bibfield  {author} {\bibinfo {author} {\bibfnamefont {S.~R.~K.}\ \bibnamefont {Rodriguez}}, \bibinfo {author} {\bibfnamefont {A.}~\bibnamefont {Amo}}, \bibinfo {author} {\bibfnamefont {I.}~\bibnamefont {Sagnes}}, \bibinfo {author} {\bibfnamefont {L.}~\bibnamefont {Le~Gratiet}}, \bibinfo {author} {\bibfnamefont {E.}~\bibnamefont {Galopin}}, \bibinfo {author} {\bibfnamefont {A.}~\bibnamefont {Lema{\^\i}tre}},\ and\ \bibinfo {author} {\bibfnamefont {J.}~\bibnamefont {Bloch}},\ }\bibfield  {title} {\bibinfo {title} {Interaction-induced hopping phase in driven-dissipative coupled photonic microcavities},\ }\href {https://doi.org/10.1038/ncomms11887} {\bibfield  {journal} {\bibinfo  {journal} {Nature Communications}\ }\textbf {\bibinfo {volume} {7}},\ \bibinfo {pages} {11887} (\bibinfo {year} {2016})}\BibitemShut {NoStop}%
\bibitem [{\citenamefont {Le~Boit\'e}\ \emph {et~al.}(2013)\citenamefont {Le~Boit\'e}, \citenamefont {Orso},\ and\ \citenamefont {Ciuti}}]{PhysRevLett.110.233601}%
  \BibitemOpen
  \bibfield  {author} {\bibinfo {author} {\bibfnamefont {A.}~\bibnamefont {Le~Boit\'e}}, \bibinfo {author} {\bibfnamefont {G.}~\bibnamefont {Orso}},\ and\ \bibinfo {author} {\bibfnamefont {C.}~\bibnamefont {Ciuti}},\ }\bibfield  {title} {\bibinfo {title} {Steady-state phases and tunneling-induced instabilities in the driven dissipative bose-hubbard model},\ }\href {https://doi.org/10.1103/PhysRevLett.110.233601} {\bibfield  {journal} {\bibinfo  {journal} {Phys. Rev. Lett.}\ }\textbf {\bibinfo {volume} {110}},\ \bibinfo {pages} {233601} (\bibinfo {year} {2013})}\BibitemShut {NoStop}%
\bibitem [{\citenamefont {Wang}\ \emph {et~al.}(2020)\citenamefont {Wang}, \citenamefont {Navarrete-Benlloch},\ and\ \citenamefont {Cai}}]{PhysRevLett.125.115301}%
  \BibitemOpen
  \bibfield  {author} {\bibinfo {author} {\bibfnamefont {Z.}~\bibnamefont {Wang}}, \bibinfo {author} {\bibfnamefont {C.}~\bibnamefont {Navarrete-Benlloch}},\ and\ \bibinfo {author} {\bibfnamefont {Z.}~\bibnamefont {Cai}},\ }\bibfield  {title} {\bibinfo {title} {Pattern formation and exotic order in driven-dissipative bose-hubbard systems},\ }\href {https://doi.org/10.1103/PhysRevLett.125.115301} {\bibfield  {journal} {\bibinfo  {journal} {Phys. Rev. Lett.}\ }\textbf {\bibinfo {volume} {125}},\ \bibinfo {pages} {115301} (\bibinfo {year} {2020})}\BibitemShut {NoStop}%
\bibitem [{\citenamefont {Zhang}\ \emph {et~al.}(2020)\citenamefont {Zhang}, \citenamefont {Yao}, \citenamefont {Feng}, \citenamefont {Hu},\ and\ \citenamefont {Chin}}]{zhang2020pattern}%
  \BibitemOpen
  \bibfield  {author} {\bibinfo {author} {\bibfnamefont {Z.}~\bibnamefont {Zhang}}, \bibinfo {author} {\bibfnamefont {K.-X.}\ \bibnamefont {Yao}}, \bibinfo {author} {\bibfnamefont {L.}~\bibnamefont {Feng}}, \bibinfo {author} {\bibfnamefont {J.}~\bibnamefont {Hu}},\ and\ \bibinfo {author} {\bibfnamefont {C.}~\bibnamefont {Chin}},\ }\bibfield  {title} {\bibinfo {title} {Pattern formation in a driven bose--einstein condensate},\ }\href {https://doi.org/10.1038/s41567-020-0839-3} {\bibfield  {journal} {\bibinfo  {journal} {Nature Physics}\ }\textbf {\bibinfo {volume} {16}},\ \bibinfo {pages} {652} (\bibinfo {year} {2020})}\BibitemShut {NoStop}%
\bibitem [{\citenamefont {Eckardt}(2017)}]{RevModPhys.89.011004}%
  \BibitemOpen
  \bibfield  {author} {\bibinfo {author} {\bibfnamefont {A.}~\bibnamefont {Eckardt}},\ }\bibfield  {title} {\bibinfo {title} {Colloquium: Atomic quantum gases in periodically driven optical lattices},\ }\href {https://doi.org/10.1103/RevModPhys.89.011004} {\bibfield  {journal} {\bibinfo  {journal} {Rev. Mod. Phys.}\ }\textbf {\bibinfo {volume} {89}},\ \bibinfo {pages} {011004} (\bibinfo {year} {2017})}\BibitemShut {NoStop}%
\bibitem [{\citenamefont {Meinert}\ \emph {et~al.}(2016)\citenamefont {Meinert}, \citenamefont {Mark}, \citenamefont {Lauber}, \citenamefont {Daley},\ and\ \citenamefont {N\"agerl}}]{PhysRevLett.116.205301}%
  \BibitemOpen
  \bibfield  {author} {\bibinfo {author} {\bibfnamefont {F.}~\bibnamefont {Meinert}}, \bibinfo {author} {\bibfnamefont {M.~J.}\ \bibnamefont {Mark}}, \bibinfo {author} {\bibfnamefont {K.}~\bibnamefont {Lauber}}, \bibinfo {author} {\bibfnamefont {A.~J.}\ \bibnamefont {Daley}},\ and\ \bibinfo {author} {\bibfnamefont {H.-C.}\ \bibnamefont {N\"agerl}},\ }\bibfield  {title} {\bibinfo {title} {Floquet engineering of correlated tunneling in the bose-hubbard model with ultracold atoms},\ }\href {https://doi.org/10.1103/PhysRevLett.116.205301} {\bibfield  {journal} {\bibinfo  {journal} {Phys. Rev. Lett.}\ }\textbf {\bibinfo {volume} {116}},\ \bibinfo {pages} {205301} (\bibinfo {year} {2016})}\BibitemShut {NoStop}%
\bibitem [{\citenamefont {Wang}\ \emph {et~al.}(2025)\citenamefont {Wang}, \citenamefont {Zhou}, \citenamefont {Shi}, \citenamefont {Huang}, \citenamefont {Yang}, \citenamefont {Zhang}, \citenamefont {Zhao}, \citenamefont {Xu}, \citenamefont {Li}, \citenamefont {Zhao}, \citenamefont {Feng}, \citenamefont {Xue}, \citenamefont {Liu}, \citenamefont {Ma}, \citenamefont {Fang}, \citenamefont {Liu}, \citenamefont {Wang}, \citenamefont {Xu}, \citenamefont {Yu}, \citenamefont {Fan},\ and\ \citenamefont {Zhao}}]{wangzt2025}%
  \BibitemOpen
  \bibfield  {author} {\bibinfo {author} {\bibfnamefont {Z.~T.}\ \bibnamefont {Wang}}, \bibinfo {author} {\bibfnamefont {S.-Y.}\ \bibnamefont {Zhou}}, \bibinfo {author} {\bibfnamefont {Y.-H.}\ \bibnamefont {Shi}}, \bibinfo {author} {\bibfnamefont {K.}~\bibnamefont {Huang}}, \bibinfo {author} {\bibfnamefont {Z.~H.}\ \bibnamefont {Yang}}, \bibinfo {author} {\bibfnamefont {J.}~\bibnamefont {Zhang}}, \bibinfo {author} {\bibfnamefont {K.}~\bibnamefont {Zhao}}, \bibinfo {author} {\bibfnamefont {Y.}~\bibnamefont {Xu}}, \bibinfo {author} {\bibfnamefont {H.}~\bibnamefont {Li}}, \bibinfo {author} {\bibfnamefont {S.~K.}\ \bibnamefont {Zhao}}, \bibinfo {author} {\bibfnamefont {Y.}~\bibnamefont {Feng}}, \bibinfo {author} {\bibfnamefont {G.}~\bibnamefont {Xue}}, \bibinfo {author} {\bibfnamefont {Y.}~\bibnamefont {Liu}}, \bibinfo {author} {\bibfnamefont {W.-G.}\ \bibnamefont {Ma}}, \bibinfo {author} {\bibfnamefont {C.-P.}\ \bibnamefont {Fang}}, \bibinfo {author} {\bibfnamefont {H.-T.}\ \bibnamefont {Liu}}, \bibinfo {author}
  {\bibfnamefont {Y.-Y.}\ \bibnamefont {Wang}}, \bibinfo {author} {\bibfnamefont {K.}~\bibnamefont {Xu}}, \bibinfo {author} {\bibfnamefont {H.}~\bibnamefont {Yu}}, \bibinfo {author} {\bibfnamefont {H.}~\bibnamefont {Fan}},\ and\ \bibinfo {author} {\bibfnamefont {S.~P.}\ \bibnamefont {Zhao}},\ }\href {https://arxiv.org/abs/2509.02180} {\bibinfo {title} {Observing two-particle correlation dynamics in tunable superconducting bose-hubbard simulators}} (\bibinfo {year} {2025}),\ \Eprint {https://arxiv.org/abs/2509.02180} {arXiv:2509.02180 [quant-ph]} \BibitemShut {NoStop}%
\bibitem [{\citenamefont {Leghtas}\ \emph {et~al.}(2015)\citenamefont {Leghtas}, \citenamefont {Touzard}, \citenamefont {Pop}, \citenamefont {Kou}, \citenamefont {Vlastakis}, \citenamefont {Petrenko}, \citenamefont {Sliwa}, \citenamefont {Narla}, \citenamefont {Shankar}, \citenamefont {Hatridge}, \citenamefont {Reagor}, \citenamefont {Frunzio}, \citenamefont {Schoelkopf}, \citenamefont {Mirrahimi},\ and\ \citenamefont {Devoret}}]{Leghtas2015}%
  \BibitemOpen
  \bibfield  {author} {\bibinfo {author} {\bibfnamefont {Z.}~\bibnamefont {Leghtas}}, \bibinfo {author} {\bibfnamefont {S.}~\bibnamefont {Touzard}}, \bibinfo {author} {\bibfnamefont {I.~M.}\ \bibnamefont {Pop}}, \bibinfo {author} {\bibfnamefont {A.}~\bibnamefont {Kou}}, \bibinfo {author} {\bibfnamefont {B.}~\bibnamefont {Vlastakis}}, \bibinfo {author} {\bibfnamefont {A.}~\bibnamefont {Petrenko}}, \bibinfo {author} {\bibfnamefont {K.~M.}\ \bibnamefont {Sliwa}}, \bibinfo {author} {\bibfnamefont {A.}~\bibnamefont {Narla}}, \bibinfo {author} {\bibfnamefont {S.}~\bibnamefont {Shankar}}, \bibinfo {author} {\bibfnamefont {M.~J.}\ \bibnamefont {Hatridge}}, \bibinfo {author} {\bibfnamefont {M.}~\bibnamefont {Reagor}}, \bibinfo {author} {\bibfnamefont {L.}~\bibnamefont {Frunzio}}, \bibinfo {author} {\bibfnamefont {R.~J.}\ \bibnamefont {Schoelkopf}}, \bibinfo {author} {\bibfnamefont {M.}~\bibnamefont {Mirrahimi}},\ and\ \bibinfo {author} {\bibfnamefont {M.~H.}\ \bibnamefont {Devoret}},\ }\bibfield  {title} {\bibinfo
  {title} {Confining the state of light to a quantum manifold by engineered two-photon loss},\ }\href {https://doi.org/10.1126/science.aaa2085} {\bibfield  {journal} {\bibinfo  {journal} {Science}\ }\textbf {\bibinfo {volume} {347}},\ \bibinfo {pages} {853} (\bibinfo {year} {2015})}\BibitemShut {NoStop}%
\bibitem [{\citenamefont {Kannan}\ \emph {et~al.}(2020)\citenamefont {Kannan}, \citenamefont {Ruckriegel}, \citenamefont {Campbell}, \citenamefont {Frisk~Kockum}, \citenamefont {Braum{\"u}ller}, \citenamefont {Kim}, \citenamefont {Kjaergaard}, \citenamefont {Krantz}, \citenamefont {Melville}, \citenamefont {Niedzielski}, \citenamefont {Veps{\"a}l{\"a}inen}, \citenamefont {Winik}, \citenamefont {Yoder}, \citenamefont {Nori}, \citenamefont {Orlando}, \citenamefont {Gustavsson},\ and\ \citenamefont {Oliver}}]{Kannan:2020aa}%
  \BibitemOpen
  \bibfield  {author} {\bibinfo {author} {\bibfnamefont {B.}~\bibnamefont {Kannan}}, \bibinfo {author} {\bibfnamefont {M.~J.}\ \bibnamefont {Ruckriegel}}, \bibinfo {author} {\bibfnamefont {D.~L.}\ \bibnamefont {Campbell}}, \bibinfo {author} {\bibfnamefont {A.}~\bibnamefont {Frisk~Kockum}}, \bibinfo {author} {\bibfnamefont {J.}~\bibnamefont {Braum{\"u}ller}}, \bibinfo {author} {\bibfnamefont {D.~K.}\ \bibnamefont {Kim}}, \bibinfo {author} {\bibfnamefont {M.}~\bibnamefont {Kjaergaard}}, \bibinfo {author} {\bibfnamefont {P.}~\bibnamefont {Krantz}}, \bibinfo {author} {\bibfnamefont {A.}~\bibnamefont {Melville}}, \bibinfo {author} {\bibfnamefont {B.~M.}\ \bibnamefont {Niedzielski}}, \bibinfo {author} {\bibfnamefont {A.}~\bibnamefont {Veps{\"a}l{\"a}inen}}, \bibinfo {author} {\bibfnamefont {R.}~\bibnamefont {Winik}}, \bibinfo {author} {\bibfnamefont {J.~L.}\ \bibnamefont {Yoder}}, \bibinfo {author} {\bibfnamefont {F.}~\bibnamefont {Nori}}, \bibinfo {author} {\bibfnamefont {T.~P.}\ \bibnamefont {Orlando}}, \bibinfo
  {author} {\bibfnamefont {S.}~\bibnamefont {Gustavsson}},\ and\ \bibinfo {author} {\bibfnamefont {W.~D.}\ \bibnamefont {Oliver}},\ }\bibfield  {title} {\bibinfo {title} {Waveguide quantum electrodynamics with superconducting artificial giant atoms},\ }\href {https://doi.org/10.1038/s41586-020-2529-9} {\bibfield  {journal} {\bibinfo  {journal} {Nature}\ }\textbf {\bibinfo {volume} {583}},\ \bibinfo {pages} {775} (\bibinfo {year} {2020})}\BibitemShut {NoStop}%
\bibitem [{\citenamefont {Vadiraj}\ \emph {et~al.}(2021)\citenamefont {Vadiraj}, \citenamefont {Ask}, \citenamefont {McConkey}, \citenamefont {Nsanzineza}, \citenamefont {Chang}, \citenamefont {Kockum},\ and\ \citenamefont {Wilson}}]{PhysRevA.103.023710}%
  \BibitemOpen
  \bibfield  {author} {\bibinfo {author} {\bibfnamefont {A.~M.}\ \bibnamefont {Vadiraj}}, \bibinfo {author} {\bibfnamefont {A.}~\bibnamefont {Ask}}, \bibinfo {author} {\bibfnamefont {T.~G.}\ \bibnamefont {McConkey}}, \bibinfo {author} {\bibfnamefont {I.}~\bibnamefont {Nsanzineza}}, \bibinfo {author} {\bibfnamefont {C.~W.~S.}\ \bibnamefont {Chang}}, \bibinfo {author} {\bibfnamefont {A.~F.}\ \bibnamefont {Kockum}},\ and\ \bibinfo {author} {\bibfnamefont {C.~M.}\ \bibnamefont {Wilson}},\ }\bibfield  {title} {\bibinfo {title} {Engineering the level structure of a giant artificial atom in waveguide quantum electrodynamics},\ }\href {https://doi.org/10.1103/PhysRevA.103.023710} {\bibfield  {journal} {\bibinfo  {journal} {Phys. Rev. A}\ }\textbf {\bibinfo {volume} {103}},\ \bibinfo {pages} {023710} (\bibinfo {year} {2021})}\BibitemShut {NoStop}%
\bibitem [{\citenamefont {Kockum}\ \emph {et~al.}(2018)\citenamefont {Kockum}, \citenamefont {Johansson},\ and\ \citenamefont {Nori}}]{PhysRevLett.120.140404}%
  \BibitemOpen
  \bibfield  {author} {\bibinfo {author} {\bibfnamefont {A.~F.}\ \bibnamefont {Kockum}}, \bibinfo {author} {\bibfnamefont {G.}~\bibnamefont {Johansson}},\ and\ \bibinfo {author} {\bibfnamefont {F.}~\bibnamefont {Nori}},\ }\bibfield  {title} {\bibinfo {title} {Decoherence-free interaction between giant atoms in waveguide quantum electrodynamics},\ }\href {https://doi.org/10.1103/PhysRevLett.120.140404} {\bibfield  {journal} {\bibinfo  {journal} {Phys. Rev. Lett.}\ }\textbf {\bibinfo {volume} {120}},\ \bibinfo {pages} {140404} (\bibinfo {year} {2018})}\BibitemShut {NoStop}%
\bibitem [{\citenamefont {Wang}\ \emph {et~al.}(2021)\citenamefont {Wang}, \citenamefont {Liu}, \citenamefont {Kockum}, \citenamefont {Li},\ and\ \citenamefont {Nori}}]{PhysRevLett.126.043602}%
  \BibitemOpen
  \bibfield  {author} {\bibinfo {author} {\bibfnamefont {X.}~\bibnamefont {Wang}}, \bibinfo {author} {\bibfnamefont {T.}~\bibnamefont {Liu}}, \bibinfo {author} {\bibfnamefont {A.~F.}\ \bibnamefont {Kockum}}, \bibinfo {author} {\bibfnamefont {H.-R.}\ \bibnamefont {Li}},\ and\ \bibinfo {author} {\bibfnamefont {F.}~\bibnamefont {Nori}},\ }\bibfield  {title} {\bibinfo {title} {Tunable chiral bound states with giant atoms},\ }\href {https://doi.org/10.1103/PhysRevLett.126.043602} {\bibfield  {journal} {\bibinfo  {journal} {Phys. Rev. Lett.}\ }\textbf {\bibinfo {volume} {126}},\ \bibinfo {pages} {043602} (\bibinfo {year} {2021})}\BibitemShut {NoStop}%
\bibitem [{\citenamefont {Deng}\ \emph {et~al.}(2002)\citenamefont {Deng}, \citenamefont {Weihs}, \citenamefont {Santori}, \citenamefont {Bloch},\ and\ \citenamefont {Yamamoto}}]{science.1074464}%
  \BibitemOpen
  \bibfield  {author} {\bibinfo {author} {\bibfnamefont {H.}~\bibnamefont {Deng}}, \bibinfo {author} {\bibfnamefont {G.}~\bibnamefont {Weihs}}, \bibinfo {author} {\bibfnamefont {C.}~\bibnamefont {Santori}}, \bibinfo {author} {\bibfnamefont {J.}~\bibnamefont {Bloch}},\ and\ \bibinfo {author} {\bibfnamefont {Y.}~\bibnamefont {Yamamoto}},\ }\bibfield  {title} {\bibinfo {title} {Condensation of semiconductor microcavity exciton polaritons},\ }\href {https://doi.org/10.1126/science.1074464} {\bibfield  {journal} {\bibinfo  {journal} {Science}\ }\textbf {\bibinfo {volume} {298}},\ \bibinfo {pages} {199} (\bibinfo {year} {2002})}\BibitemShut {NoStop}%
\bibitem [{\citenamefont {Balili}\ \emph {et~al.}(2007)\citenamefont {Balili}, \citenamefont {Hartwell}, \citenamefont {Snoke}, \citenamefont {Pfeiffer},\ and\ \citenamefont {West}}]{science.1140990}%
  \BibitemOpen
  \bibfield  {author} {\bibinfo {author} {\bibfnamefont {R.}~\bibnamefont {Balili}}, \bibinfo {author} {\bibfnamefont {V.}~\bibnamefont {Hartwell}}, \bibinfo {author} {\bibfnamefont {D.}~\bibnamefont {Snoke}}, \bibinfo {author} {\bibfnamefont {L.}~\bibnamefont {Pfeiffer}},\ and\ \bibinfo {author} {\bibfnamefont {K.}~\bibnamefont {West}},\ }\bibfield  {title} {\bibinfo {title} {Bose-einstein condensation of microcavity polaritons in a trap},\ }\href {https://doi.org/10.1126/science.1140990} {\bibfield  {journal} {\bibinfo  {journal} {Science}\ }\textbf {\bibinfo {volume} {316}},\ \bibinfo {pages} {1007} (\bibinfo {year} {2007})}\BibitemShut {NoStop}%
\bibitem [{\citenamefont {Hartmann}(2016)}]{Hartmann2016}%
  \BibitemOpen
  \bibfield  {author} {\bibinfo {author} {\bibfnamefont {M.~J.}\ \bibnamefont {Hartmann}},\ }\bibfield  {title} {\bibinfo {title} {Quantum simulation with interacting photons},\ }\href@noop {} {\bibfield  {journal} {\bibinfo  {journal} {Journal of Optics}\ }\textbf {\bibinfo {volume} {18}},\ \bibinfo {pages} {104005} (\bibinfo {year} {2016})}\BibitemShut {NoStop}%
\bibitem [{\citenamefont {Ebadi}\ \emph {et~al.}(2021)\citenamefont {Ebadi}, \citenamefont {Wang}, \citenamefont {Levine}, \citenamefont {Keesling}, \citenamefont {Semeghini}, \citenamefont {Omran}, \citenamefont {Bluvstein}, \citenamefont {Samajdar}, \citenamefont {Pichler}, \citenamefont {Ho}, \citenamefont {Choi}, \citenamefont {Sachdev}, \citenamefont {Greiner}, \citenamefont {Vuleti{\'c}},\ and\ \citenamefont {Lukin}}]{Ebadi:2021aa}%
  \BibitemOpen
  \bibfield  {author} {\bibinfo {author} {\bibfnamefont {S.}~\bibnamefont {Ebadi}}, \bibinfo {author} {\bibfnamefont {T.~T.}\ \bibnamefont {Wang}}, \bibinfo {author} {\bibfnamefont {H.}~\bibnamefont {Levine}}, \bibinfo {author} {\bibfnamefont {A.}~\bibnamefont {Keesling}}, \bibinfo {author} {\bibfnamefont {G.}~\bibnamefont {Semeghini}}, \bibinfo {author} {\bibfnamefont {A.}~\bibnamefont {Omran}}, \bibinfo {author} {\bibfnamefont {D.}~\bibnamefont {Bluvstein}}, \bibinfo {author} {\bibfnamefont {R.}~\bibnamefont {Samajdar}}, \bibinfo {author} {\bibfnamefont {H.}~\bibnamefont {Pichler}}, \bibinfo {author} {\bibfnamefont {W.~W.}\ \bibnamefont {Ho}}, \bibinfo {author} {\bibfnamefont {S.}~\bibnamefont {Choi}}, \bibinfo {author} {\bibfnamefont {S.}~\bibnamefont {Sachdev}}, \bibinfo {author} {\bibfnamefont {M.}~\bibnamefont {Greiner}}, \bibinfo {author} {\bibfnamefont {V.}~\bibnamefont {Vuleti{\'c}}},\ and\ \bibinfo {author} {\bibfnamefont {M.~D.}\ \bibnamefont {Lukin}},\ }\bibfield  {title} {\bibinfo {title}
  {Quantum phases of matter on a 256-atom programmable quantum simulator},\ }\href {https://doi.org/10.1038/s41586-021-03582-4} {\bibfield  {journal} {\bibinfo  {journal} {Nature}\ }\textbf {\bibinfo {volume} {595}},\ \bibinfo {pages} {227} (\bibinfo {year} {2021})}\BibitemShut {NoStop}%
\bibitem [{\citenamefont {Browaeys}\ and\ \citenamefont {Lahaye}(2020)}]{Browaeys:2020aa}%
  \BibitemOpen
  \bibfield  {author} {\bibinfo {author} {\bibfnamefont {A.}~\bibnamefont {Browaeys}}\ and\ \bibinfo {author} {\bibfnamefont {T.}~\bibnamefont {Lahaye}},\ }\bibfield  {title} {\bibinfo {title} {Many-body physics with individually controlled rydberg atoms},\ }\href {https://doi.org/10.1038/s41567-019-0733-z} {\bibfield  {journal} {\bibinfo  {journal} {Nature Physics}\ }\textbf {\bibinfo {volume} {16}},\ \bibinfo {pages} {132} (\bibinfo {year} {2020})}\BibitemShut {NoStop}%
\bibitem [{\citenamefont {Blais}\ \emph {et~al.}(2021)\citenamefont {Blais}, \citenamefont {Grimsmo}, \citenamefont {Girvin},\ and\ \citenamefont {Wallraff}}]{RevModPhys.93.025005}%
  \BibitemOpen
  \bibfield  {author} {\bibinfo {author} {\bibfnamefont {A.}~\bibnamefont {Blais}}, \bibinfo {author} {\bibfnamefont {A.~L.}\ \bibnamefont {Grimsmo}}, \bibinfo {author} {\bibfnamefont {S.~M.}\ \bibnamefont {Girvin}},\ and\ \bibinfo {author} {\bibfnamefont {A.}~\bibnamefont {Wallraff}},\ }\bibfield  {title} {\bibinfo {title} {Circuit quantum electrodynamics},\ }\href {https://doi.org/10.1103/RevModPhys.93.025005} {\bibfield  {journal} {\bibinfo  {journal} {Rev. Mod. Phys.}\ }\textbf {\bibinfo {volume} {93}},\ \bibinfo {pages} {025005} (\bibinfo {year} {2021})}\BibitemShut {NoStop}%
\bibitem [{\citenamefont {Ma}\ \emph {et~al.}(2019)\citenamefont {Ma}, \citenamefont {Saxberg}, \citenamefont {Owens}, \citenamefont {Leung}, \citenamefont {Lu}, \citenamefont {Simon},\ and\ \citenamefont {Schuster}}]{Ma:2019aa}%
  \BibitemOpen
  \bibfield  {author} {\bibinfo {author} {\bibfnamefont {R.}~\bibnamefont {Ma}}, \bibinfo {author} {\bibfnamefont {B.}~\bibnamefont {Saxberg}}, \bibinfo {author} {\bibfnamefont {C.}~\bibnamefont {Owens}}, \bibinfo {author} {\bibfnamefont {N.}~\bibnamefont {Leung}}, \bibinfo {author} {\bibfnamefont {Y.}~\bibnamefont {Lu}}, \bibinfo {author} {\bibfnamefont {J.}~\bibnamefont {Simon}},\ and\ \bibinfo {author} {\bibfnamefont {D.~I.}\ \bibnamefont {Schuster}},\ }\bibfield  {title} {\bibinfo {title} {A dissipatively stabilized mott insulator of photons},\ }\href {https://doi.org/10.1038/s41586-019-0897-9} {\bibfield  {journal} {\bibinfo  {journal} {Nature}\ }\textbf {\bibinfo {volume} {566}},\ \bibinfo {pages} {51} (\bibinfo {year} {2019})}\BibitemShut {NoStop}%
\bibitem [{\citenamefont {Klembt}\ \emph {et~al.}(2018)\citenamefont {Klembt}, \citenamefont {Harder}, \citenamefont {Egorov}, \citenamefont {Winkler}, \citenamefont {Ge}, \citenamefont {Bandres}, \citenamefont {Emmerling}, \citenamefont {Worschech}, \citenamefont {Liew}, \citenamefont {Segev}, \citenamefont {Schneider},\ and\ \citenamefont {H{\"o}fling}}]{Klembt:2018aa}%
  \BibitemOpen
  \bibfield  {author} {\bibinfo {author} {\bibfnamefont {S.}~\bibnamefont {Klembt}}, \bibinfo {author} {\bibfnamefont {T.~H.}\ \bibnamefont {Harder}}, \bibinfo {author} {\bibfnamefont {O.~A.}\ \bibnamefont {Egorov}}, \bibinfo {author} {\bibfnamefont {K.}~\bibnamefont {Winkler}}, \bibinfo {author} {\bibfnamefont {R.}~\bibnamefont {Ge}}, \bibinfo {author} {\bibfnamefont {M.~A.}\ \bibnamefont {Bandres}}, \bibinfo {author} {\bibfnamefont {M.}~\bibnamefont {Emmerling}}, \bibinfo {author} {\bibfnamefont {L.}~\bibnamefont {Worschech}}, \bibinfo {author} {\bibfnamefont {T.~C.~H.}\ \bibnamefont {Liew}}, \bibinfo {author} {\bibfnamefont {M.}~\bibnamefont {Segev}}, \bibinfo {author} {\bibfnamefont {C.}~\bibnamefont {Schneider}},\ and\ \bibinfo {author} {\bibfnamefont {S.}~\bibnamefont {H{\"o}fling}},\ }\bibfield  {title} {\bibinfo {title} {Exciton-polariton topological insulator},\ }\href {https://doi.org/10.1038/s41586-018-0601-5} {\bibfield  {journal} {\bibinfo  {journal} {Nature}\ }\textbf {\bibinfo {volume} {562}},\
  \bibinfo {pages} {552} (\bibinfo {year} {2018})}\BibitemShut {NoStop}%
\bibitem [{\citenamefont {Carusotto}\ \emph {et~al.}(2020)\citenamefont {Carusotto}, \citenamefont {Houck}, \citenamefont {Koll{\'a}r}, \citenamefont {Roushan}, \citenamefont {Schuster},\ and\ \citenamefont {Simon}}]{Carusotto:2020aa}%
  \BibitemOpen
  \bibfield  {author} {\bibinfo {author} {\bibfnamefont {I.}~\bibnamefont {Carusotto}}, \bibinfo {author} {\bibfnamefont {A.~A.}\ \bibnamefont {Houck}}, \bibinfo {author} {\bibfnamefont {A.~J.}\ \bibnamefont {Koll{\'a}r}}, \bibinfo {author} {\bibfnamefont {P.}~\bibnamefont {Roushan}}, \bibinfo {author} {\bibfnamefont {D.~I.}\ \bibnamefont {Schuster}},\ and\ \bibinfo {author} {\bibfnamefont {J.}~\bibnamefont {Simon}},\ }\bibfield  {title} {\bibinfo {title} {Photonic materials in circuit quantum electrodynamics},\ }\href {https://doi.org/10.1038/s41567-020-0815-y} {\bibfield  {journal} {\bibinfo  {journal} {Nature Physics}\ }\textbf {\bibinfo {volume} {16}},\ \bibinfo {pages} {268} (\bibinfo {year} {2020})}\BibitemShut {NoStop}%
\bibitem [{\citenamefont {Monroe}\ \emph {et~al.}(2021)\citenamefont {Monroe}, \citenamefont {Campbell}, \citenamefont {Duan}, \citenamefont {Gong}, \citenamefont {Gorshkov}, \citenamefont {Hess}, \citenamefont {Islam}, \citenamefont {Kim}, \citenamefont {Linke}, \citenamefont {Pagano}, \citenamefont {Richerme}, \citenamefont {Senko},\ and\ \citenamefont {Yao}}]{RevModPhys.93.025001}%
  \BibitemOpen
  \bibfield  {author} {\bibinfo {author} {\bibfnamefont {C.}~\bibnamefont {Monroe}}, \bibinfo {author} {\bibfnamefont {W.~C.}\ \bibnamefont {Campbell}}, \bibinfo {author} {\bibfnamefont {L.-M.}\ \bibnamefont {Duan}}, \bibinfo {author} {\bibfnamefont {Z.-X.}\ \bibnamefont {Gong}}, \bibinfo {author} {\bibfnamefont {A.~V.}\ \bibnamefont {Gorshkov}}, \bibinfo {author} {\bibfnamefont {P.~W.}\ \bibnamefont {Hess}}, \bibinfo {author} {\bibfnamefont {R.}~\bibnamefont {Islam}}, \bibinfo {author} {\bibfnamefont {K.}~\bibnamefont {Kim}}, \bibinfo {author} {\bibfnamefont {N.~M.}\ \bibnamefont {Linke}}, \bibinfo {author} {\bibfnamefont {G.}~\bibnamefont {Pagano}}, \bibinfo {author} {\bibfnamefont {P.}~\bibnamefont {Richerme}}, \bibinfo {author} {\bibfnamefont {C.}~\bibnamefont {Senko}},\ and\ \bibinfo {author} {\bibfnamefont {N.~Y.}\ \bibnamefont {Yao}},\ }\bibfield  {title} {\bibinfo {title} {Programmable quantum simulations of spin systems with trapped ions},\ }\href {https://doi.org/10.1103/RevModPhys.93.025001}
  {\bibfield  {journal} {\bibinfo  {journal} {Rev. Mod. Phys.}\ }\textbf {\bibinfo {volume} {93}},\ \bibinfo {pages} {025001} (\bibinfo {year} {2021})}\BibitemShut {NoStop}%
\bibitem [{\citenamefont {Gross}\ and\ \citenamefont {Bloch}(2017)}]{GrossBloch2017}%
  \BibitemOpen
  \bibfield  {author} {\bibinfo {author} {\bibfnamefont {C.}~\bibnamefont {Gross}}\ and\ \bibinfo {author} {\bibfnamefont {I.}~\bibnamefont {Bloch}},\ }\bibfield  {title} {\bibinfo {title} {Quantum simulations with ultracold atoms in optical lattices},\ }\href {https://doi.org/10.1126/science.aal3837} {\bibfield  {journal} {\bibinfo  {journal} {Science}\ }\textbf {\bibinfo {volume} {357}},\ \bibinfo {pages} {995} (\bibinfo {year} {2017})}\BibitemShut {NoStop}%
\bibitem [{\citenamefont {Barontini}\ \emph {et~al.}(2013)\citenamefont {Barontini}, \citenamefont {Labouvie}, \citenamefont {Stubenrauch}, \citenamefont {Vogler}, \citenamefont {Guarrera},\ and\ \citenamefont {Ott}}]{PhysRevLett.110.035302}%
  \BibitemOpen
  \bibfield  {author} {\bibinfo {author} {\bibfnamefont {G.}~\bibnamefont {Barontini}}, \bibinfo {author} {\bibfnamefont {R.}~\bibnamefont {Labouvie}}, \bibinfo {author} {\bibfnamefont {F.}~\bibnamefont {Stubenrauch}}, \bibinfo {author} {\bibfnamefont {A.}~\bibnamefont {Vogler}}, \bibinfo {author} {\bibfnamefont {V.}~\bibnamefont {Guarrera}},\ and\ \bibinfo {author} {\bibfnamefont {H.}~\bibnamefont {Ott}},\ }\bibfield  {title} {\bibinfo {title} {Controlling the dynamics of an open many-body quantum system with localized dissipation},\ }\href {https://doi.org/10.1103/PhysRevLett.110.035302} {\bibfield  {journal} {\bibinfo  {journal} {Phys. Rev. Lett.}\ }\textbf {\bibinfo {volume} {110}},\ \bibinfo {pages} {035302} (\bibinfo {year} {2013})}\BibitemShut {NoStop}%
\bibitem [{\citenamefont {Kosior}\ \emph {et~al.}(2024)\citenamefont {Kosior}, \citenamefont {Gietka}, \citenamefont {Mivehvar},\ and\ \citenamefont {Ritsch}}]{kosior2024nonequilibrium}%
  \BibitemOpen
  \bibfield  {author} {\bibinfo {author} {\bibfnamefont {A.}~\bibnamefont {Kosior}}, \bibinfo {author} {\bibfnamefont {K.}~\bibnamefont {Gietka}}, \bibinfo {author} {\bibfnamefont {F.}~\bibnamefont {Mivehvar}},\ and\ \bibinfo {author} {\bibfnamefont {H.}~\bibnamefont {Ritsch}},\ }\bibfield  {title} {\bibinfo {title} {Nonequilibrium nonlinear effects and dynamical boson condensation in a driven-dissipative wannier-stark lattice},\ }\href {https://doi.org/10.1103/PhysRevB.110.L100303} {\bibfield  {journal} {\bibinfo  {journal} {Phys. Rev. B}\ }\textbf {\bibinfo {volume} {110}},\ \bibinfo {pages} {L100303} (\bibinfo {year} {2024})}\BibitemShut {NoStop}%
\bibitem [{\citenamefont {Labouvie}\ \emph {et~al.}(2015)\citenamefont {Labouvie}, \citenamefont {Santra}, \citenamefont {Heun}, \citenamefont {Wimberger},\ and\ \citenamefont {Ott}}]{PhysRevLett.115.050601}%
  \BibitemOpen
  \bibfield  {author} {\bibinfo {author} {\bibfnamefont {R.}~\bibnamefont {Labouvie}}, \bibinfo {author} {\bibfnamefont {B.}~\bibnamefont {Santra}}, \bibinfo {author} {\bibfnamefont {S.}~\bibnamefont {Heun}}, \bibinfo {author} {\bibfnamefont {S.}~\bibnamefont {Wimberger}},\ and\ \bibinfo {author} {\bibfnamefont {H.}~\bibnamefont {Ott}},\ }\bibfield  {title} {\bibinfo {title} {Negative differential conductivity in an interacting quantum gas},\ }\href {https://doi.org/10.1103/PhysRevLett.115.050601} {\bibfield  {journal} {\bibinfo  {journal} {Phys. Rev. Lett.}\ }\textbf {\bibinfo {volume} {115}},\ \bibinfo {pages} {050601} (\bibinfo {year} {2015})}\BibitemShut {NoStop}%
\bibitem [{\citenamefont {Li}\ \emph {et~al.}(2022)\citenamefont {Li}, \citenamefont {Claude}, \citenamefont {Boulier}, \citenamefont {Giacobino}, \citenamefont {Glorieux}, \citenamefont {Bramati},\ and\ \citenamefont {Ciuti}}]{PhysRevLett.128.093601}%
  \BibitemOpen
  \bibfield  {author} {\bibinfo {author} {\bibfnamefont {Z.}~\bibnamefont {Li}}, \bibinfo {author} {\bibfnamefont {F.}~\bibnamefont {Claude}}, \bibinfo {author} {\bibfnamefont {T.}~\bibnamefont {Boulier}}, \bibinfo {author} {\bibfnamefont {E.}~\bibnamefont {Giacobino}}, \bibinfo {author} {\bibfnamefont {Q.}~\bibnamefont {Glorieux}}, \bibinfo {author} {\bibfnamefont {A.}~\bibnamefont {Bramati}},\ and\ \bibinfo {author} {\bibfnamefont {C.}~\bibnamefont {Ciuti}},\ }\bibfield  {title} {\bibinfo {title} {Dissipative phase transition with driving-controlled spatial dimension and diffusive boundary conditions},\ }\href {https://doi.org/10.1103/PhysRevLett.128.093601} {\bibfield  {journal} {\bibinfo  {journal} {Phys. Rev. Lett.}\ }\textbf {\bibinfo {volume} {128}},\ \bibinfo {pages} {093601} (\bibinfo {year} {2022})}\BibitemShut {NoStop}%
\bibitem [{\citenamefont {Sinatra}\ \emph {et~al.}(2002)\citenamefont {Sinatra}, \citenamefont {Lobo},\ and\ \citenamefont {Castin}}]{Alice_Sinatra_2002}%
  \BibitemOpen
  \bibfield  {author} {\bibinfo {author} {\bibfnamefont {A.}~\bibnamefont {Sinatra}}, \bibinfo {author} {\bibfnamefont {C.}~\bibnamefont {Lobo}},\ and\ \bibinfo {author} {\bibfnamefont {Y.}~\bibnamefont {Castin}},\ }\bibfield  {title} {\bibinfo {title} {The truncated wigner method for bose-condensed gases: limits of validity and applications1},\ }\href {https://doi.org/10.1088/0953-4075/35/17/301} {\bibfield  {journal} {\bibinfo  {journal} {Journal of Physics B: Atomic, Molecular and Optical Physics}\ }\textbf {\bibinfo {volume} {35}},\ \bibinfo {pages} {3599} (\bibinfo {year} {2002})}\BibitemShut {NoStop}%
\bibitem [{\citenamefont {Carusotto}\ and\ \citenamefont {Ciuti}(2013)}]{Carusotto2013}%
  \BibitemOpen
  \bibfield  {author} {\bibinfo {author} {\bibfnamefont {I.}~\bibnamefont {Carusotto}}\ and\ \bibinfo {author} {\bibfnamefont {C.}~\bibnamefont {Ciuti}},\ }\bibfield  {title} {\bibinfo {title} {Quantum fluids of light},\ }\href {https://doi.org/10.1103/RevModPhys.85.299} {\bibfield  {journal} {\bibinfo  {journal} {Rev. Mod. Phys.}\ }\textbf {\bibinfo {volume} {85}},\ \bibinfo {pages} {299} (\bibinfo {year} {2013})}\BibitemShut {NoStop}%
\bibitem [{\citenamefont {Provazza}\ and\ \citenamefont {Tempelaar}(2022)}]{PhysRevA.106.042406}%
  \BibitemOpen
  \bibfield  {author} {\bibinfo {author} {\bibfnamefont {J.}~\bibnamefont {Provazza}}\ and\ \bibinfo {author} {\bibfnamefont {R.}~\bibnamefont {Tempelaar}},\ }\bibfield  {title} {\bibinfo {title} {Perturbation theory under the truncated wigner approximation: How system-environment entanglement formation drives quantum decoherence},\ }\href {https://doi.org/10.1103/PhysRevA.106.042406} {\bibfield  {journal} {\bibinfo  {journal} {Phys. Rev. A}\ }\textbf {\bibinfo {volume} {106}},\ \bibinfo {pages} {042406} (\bibinfo {year} {2022})}\BibitemShut {NoStop}%
\bibitem [{\citenamefont {Sundar}\ \emph {et~al.}(2019)\citenamefont {Sundar}, \citenamefont {Wang},\ and\ \citenamefont {Hazzard}}]{PhysRevA.99.043627}%
  \BibitemOpen
  \bibfield  {author} {\bibinfo {author} {\bibfnamefont {B.}~\bibnamefont {Sundar}}, \bibinfo {author} {\bibfnamefont {K.~C.}\ \bibnamefont {Wang}},\ and\ \bibinfo {author} {\bibfnamefont {K.~R.~A.}\ \bibnamefont {Hazzard}},\ }\bibfield  {title} {\bibinfo {title} {Analysis of continuous and discrete wigner approximations for spin dynamics},\ }\href {https://doi.org/10.1103/PhysRevA.99.043627} {\bibfield  {journal} {\bibinfo  {journal} {Phys. Rev. A}\ }\textbf {\bibinfo {volume} {99}},\ \bibinfo {pages} {043627} (\bibinfo {year} {2019})}\BibitemShut {NoStop}%
\bibitem [{sup()}]{supplementary}%
  \BibitemOpen
  \href@noop {} {}\bibinfo {note} {See Supplemental Material for the detailed derivation process and more numerical results.}\BibitemShut {Stop}%
\bibitem [{\citenamefont {Lugiato}\ and\ \citenamefont {Lefever}(1987)}]{PhysRevLett.58.2209}%
  \BibitemOpen
  \bibfield  {author} {\bibinfo {author} {\bibfnamefont {L.~A.}\ \bibnamefont {Lugiato}}\ and\ \bibinfo {author} {\bibfnamefont {R.}~\bibnamefont {Lefever}},\ }\bibfield  {title} {\bibinfo {title} {Spatial dissipative structures in passive optical systems},\ }\href {https://doi.org/10.1103/PhysRevLett.58.2209} {\bibfield  {journal} {\bibinfo  {journal} {Phys. Rev. Lett.}\ }\textbf {\bibinfo {volume} {58}},\ \bibinfo {pages} {2209} (\bibinfo {year} {1987})}\BibitemShut {NoStop}%
\bibitem [{\citenamefont {Gladilin}\ \emph {et~al.}(2014)\citenamefont {Gladilin}, \citenamefont {Ji},\ and\ \citenamefont {Wouters}}]{PhysRevA.90.023615}%
  \BibitemOpen
  \bibfield  {author} {\bibinfo {author} {\bibfnamefont {V.~N.}\ \bibnamefont {Gladilin}}, \bibinfo {author} {\bibfnamefont {K.}~\bibnamefont {Ji}},\ and\ \bibinfo {author} {\bibfnamefont {M.}~\bibnamefont {Wouters}},\ }\bibfield  {title} {\bibinfo {title} {Spatial coherence of weakly interacting one-dimensional nonequilibrium bosonic quantum fluids},\ }\href {https://doi.org/10.1103/PhysRevA.90.023615} {\bibfield  {journal} {\bibinfo  {journal} {Phys. Rev. A}\ }\textbf {\bibinfo {volume} {90}},\ \bibinfo {pages} {023615} (\bibinfo {year} {2014})}\BibitemShut {NoStop}%
\bibitem [{\citenamefont {Ji}\ \emph {et~al.}(2015)\citenamefont {Ji}, \citenamefont {Gladilin},\ and\ \citenamefont {Wouters}}]{PhysRevB.91.045301}%
  \BibitemOpen
  \bibfield  {author} {\bibinfo {author} {\bibfnamefont {K.}~\bibnamefont {Ji}}, \bibinfo {author} {\bibfnamefont {V.~N.}\ \bibnamefont {Gladilin}},\ and\ \bibinfo {author} {\bibfnamefont {M.}~\bibnamefont {Wouters}},\ }\bibfield  {title} {\bibinfo {title} {Temporal coherence of one-dimensional nonequilibrium quantum fluids},\ }\href {https://doi.org/10.1103/PhysRevB.91.045301} {\bibfield  {journal} {\bibinfo  {journal} {Phys. Rev. B}\ }\textbf {\bibinfo {volume} {91}},\ \bibinfo {pages} {045301} (\bibinfo {year} {2015})}\BibitemShut {NoStop}%
\bibitem [{\citenamefont {Sieberer}\ \emph {et~al.}(2025)\citenamefont {Sieberer}, \citenamefont {Buchhold}, \citenamefont {Marino},\ and\ \citenamefont {Diehl}}]{RevModPhys.97.025004}%
  \BibitemOpen
  \bibfield  {author} {\bibinfo {author} {\bibfnamefont {L.~M.}\ \bibnamefont {Sieberer}}, \bibinfo {author} {\bibfnamefont {M.}~\bibnamefont {Buchhold}}, \bibinfo {author} {\bibfnamefont {J.}~\bibnamefont {Marino}},\ and\ \bibinfo {author} {\bibfnamefont {S.}~\bibnamefont {Diehl}},\ }\bibfield  {title} {\bibinfo {title} {Universality in driven open quantum matter},\ }\href {https://doi.org/10.1103/RevModPhys.97.025004} {\bibfield  {journal} {\bibinfo  {journal} {Rev. Mod. Phys.}\ }\textbf {\bibinfo {volume} {97}},\ \bibinfo {pages} {025004} (\bibinfo {year} {2025})}\BibitemShut {NoStop}%
\bibitem [{\citenamefont {Ahufinger}\ \emph {et~al.}(2004)\citenamefont {Ahufinger}, \citenamefont {Sanpera}, \citenamefont {Pedri}, \citenamefont {Santos},\ and\ \citenamefont {Lewenstein}}]{PhysRevA.69.053604}%
  \BibitemOpen
  \bibfield  {author} {\bibinfo {author} {\bibfnamefont {V.}~\bibnamefont {Ahufinger}}, \bibinfo {author} {\bibfnamefont {A.}~\bibnamefont {Sanpera}}, \bibinfo {author} {\bibfnamefont {P.}~\bibnamefont {Pedri}}, \bibinfo {author} {\bibfnamefont {L.}~\bibnamefont {Santos}},\ and\ \bibinfo {author} {\bibfnamefont {M.}~\bibnamefont {Lewenstein}},\ }\bibfield  {title} {\bibinfo {title} {Creation and mobility of discrete solitons in bose-einstein condensates},\ }\href {https://doi.org/10.1103/PhysRevA.69.053604} {\bibfield  {journal} {\bibinfo  {journal} {Phys. Rev. A}\ }\textbf {\bibinfo {volume} {69}},\ \bibinfo {pages} {053604} (\bibinfo {year} {2004})}\BibitemShut {NoStop}%
\bibitem [{\citenamefont {Jenkinson}\ and\ \citenamefont {Weinstein}(2017)}]{Jenkinson:2017aa}%
  \BibitemOpen
  \bibfield  {author} {\bibinfo {author} {\bibfnamefont {M.}~\bibnamefont {Jenkinson}}\ and\ \bibinfo {author} {\bibfnamefont {M.~I.}\ \bibnamefont {Weinstein}},\ }\bibfield  {title} {\bibinfo {title} {Discrete solitary waves in systems with nonlocal interactions and the peierls--nabarro barrier},\ }\href {https://doi.org/10.1007/s00220-017-2839-4} {\bibfield  {journal} {\bibinfo  {journal} {Communications in Mathematical Physics}\ }\textbf {\bibinfo {volume} {351}},\ \bibinfo {pages} {45} (\bibinfo {year} {2017})}\BibitemShut {NoStop}%
\bibitem [{\citenamefont {Molignini}(2025)}]{szdc-61nl}%
  \BibitemOpen
  \bibfield  {author} {\bibinfo {author} {\bibfnamefont {P.}~\bibnamefont {Molignini}},\ }\bibfield  {title} {\bibinfo {title} {Stability of quasicrystalline ultracold fermions to dipolar interactions},\ }\href {https://doi.org/10.1103/szdc-61nl} {\bibfield  {journal} {\bibinfo  {journal} {Phys. Rev. Res.}\ }\textbf {\bibinfo {volume} {7}},\ \bibinfo {pages} {L032026} (\bibinfo {year} {2025})}\BibitemShut {NoStop}%
\bibitem [{\citenamefont {M{\'a}k}\ \emph {et~al.}(2024)\citenamefont {M{\'a}k}, \citenamefont {Bhaseen},\ and\ \citenamefont {Pal}}]{Mak:2024aa}%
  \BibitemOpen
  \bibfield  {author} {\bibinfo {author} {\bibfnamefont {J.}~\bibnamefont {M{\'a}k}}, \bibinfo {author} {\bibfnamefont {M.~J.}\ \bibnamefont {Bhaseen}},\ and\ \bibinfo {author} {\bibfnamefont {A.}~\bibnamefont {Pal}},\ }\bibfield  {title} {\bibinfo {title} {Statics and dynamics of non-hermitian many-body localization},\ }\href {https://doi.org/10.1038/s42005-024-01576-y} {\bibfield  {journal} {\bibinfo  {journal} {Communications Physics}\ }\textbf {\bibinfo {volume} {7}},\ \bibinfo {pages} {92} (\bibinfo {year} {2024})}\BibitemShut {NoStop}%
\bibitem [{\citenamefont {Schreiber}\ \emph {et~al.}(2015)\citenamefont {Schreiber}, \citenamefont {Hodgman}, \citenamefont {Bordia}, \citenamefont {L{\"u}schen}, \citenamefont {Fischer}, \citenamefont {Vosk}, \citenamefont {Altman}, \citenamefont {Schneider},\ and\ \citenamefont {Bloch}}]{Schreiber2015}%
  \BibitemOpen
  \bibfield  {author} {\bibinfo {author} {\bibfnamefont {M.}~\bibnamefont {Schreiber}}, \bibinfo {author} {\bibfnamefont {S.~S.}\ \bibnamefont {Hodgman}}, \bibinfo {author} {\bibfnamefont {P.}~\bibnamefont {Bordia}}, \bibinfo {author} {\bibfnamefont {H.~P.}\ \bibnamefont {L{\"u}schen}}, \bibinfo {author} {\bibfnamefont {M.~H.}\ \bibnamefont {Fischer}}, \bibinfo {author} {\bibfnamefont {R.}~\bibnamefont {Vosk}}, \bibinfo {author} {\bibfnamefont {E.}~\bibnamefont {Altman}}, \bibinfo {author} {\bibfnamefont {U.}~\bibnamefont {Schneider}},\ and\ \bibinfo {author} {\bibfnamefont {I.}~\bibnamefont {Bloch}},\ }\bibfield  {title} {\bibinfo {title} {Observation of many-body localization of interacting fermions in a quasirandom optical lattice},\ }\href {https://doi.org/10.1126/science.aaa7432} {\bibfield  {journal} {\bibinfo  {journal} {Science}\ }\textbf {\bibinfo {volume} {349}},\ \bibinfo {pages} {842} (\bibinfo {year} {2015})}\BibitemShut {NoStop}%
\bibitem [{\citenamefont {L\"uschen}\ \emph {et~al.}(2017)\citenamefont {L\"uschen}, \citenamefont {Bordia}, \citenamefont {Scherg}, \citenamefont {Alet}, \citenamefont {Altman}, \citenamefont {Schneider},\ and\ \citenamefont {Bloch}}]{PhysRevLett.119.260401}%
  \BibitemOpen
  \bibfield  {author} {\bibinfo {author} {\bibfnamefont {H.~P.}\ \bibnamefont {L\"uschen}}, \bibinfo {author} {\bibfnamefont {P.}~\bibnamefont {Bordia}}, \bibinfo {author} {\bibfnamefont {S.}~\bibnamefont {Scherg}}, \bibinfo {author} {\bibfnamefont {F.}~\bibnamefont {Alet}}, \bibinfo {author} {\bibfnamefont {E.}~\bibnamefont {Altman}}, \bibinfo {author} {\bibfnamefont {U.}~\bibnamefont {Schneider}},\ and\ \bibinfo {author} {\bibfnamefont {I.}~\bibnamefont {Bloch}},\ }\bibfield  {title} {\bibinfo {title} {Observation of slow dynamics near the many-body localization transition in one-dimensional quasiperiodic systems},\ }\href {https://doi.org/10.1103/PhysRevLett.119.260401} {\bibfield  {journal} {\bibinfo  {journal} {Phys. Rev. Lett.}\ }\textbf {\bibinfo {volume} {119}},\ \bibinfo {pages} {260401} (\bibinfo {year} {2017})}\BibitemShut {NoStop}%
\end{thebibliography}
\end{document}

\renewcommand{\thefigure}{S\arabic{figure}}
\setcounter{figure}{0}
\renewcommand{\theequation}{S\arabic{equation}} 
\setcounter{equation}{0}

% \preprint{APS/123-QED}

\title{Supplemental Material for:\\ ``Dissipation-Selected Resonant Fronts in a Driven-Dissipative Bose-Hubbard Lattice''}

\author{Wei-Guo Ma}
\email{weiguo.m@iphy.ac.cn}
\affiliation{Beijing National Laboratory for Condensed Matter Physics,
Institute of Physics, Chinese Academy of Sciences, Beijing 100190, China}
\affiliation{School of Physical Sciences, University of Chinese Academy of Sciences, Beijing 100049, China}

\author{Heng Fan}
\email{hfan@iphy.ac.cn}
\affiliation{Beijing National Laboratory for Condensed Matter Physics,
Institute of Physics, Chinese Academy of Sciences, Beijing 100190, China}
\affiliation{School of Physical Sciences, University of Chinese Academy of Sciences, Beijing 100049, China}
\affiliation{Beijing Key Laboratory of Advanced Quantum Technology,
Beijing Academy of Quantum Information Sciences, Beijing 100193, China}
\affiliation{Hefei National Laboratory, Hefei 230088, China}
\affiliation{Songshan Lake Materials Laboratory, Dongguan, Guangdong 523808, China}

\date{\today}

\maketitle

\section{I. Truncated Wigner approximation for driven-dissipative dynamics}
\label{sec:SM_TWA}

The generalized Gross-Pitaevskii equation (gGPE) captures large-scale condensate dynamics but neglects quantum fluctuations. To include leading-order fluctuation effects while remaining tractable on large two-dimensional lattices, we employ the truncated Wigner approximation (TWA), which has proven effective for driven-dissipative bosonic platforms when mode occupations are sufficiently large~\cite{Alice_Sinatra_2002, PhysRevB.79.165302, PhysRevApplied.22.054061}. In the initial-value formulation of TWA (IV-TWA)~\cite{Alice_Sinatra_2002, PhysRevA.106.042406, PhysRevA.99.043627}, each stochastic realization is evolved according to the gGPE,
\begin{equation}
  i\frac{d\alpha_j}{dt}
  = \Big[\Delta\omega + F x_j - \tfrac{i}{2}(\kappa_0+\gamma x_j)\Big]\alpha_j
  + U|\alpha_j|^2\alpha_j
  - J\sum_{k\in\langle j\rangle}\alpha_k
  + \Omega.
  \label{eq:S_TWA_eom}
\end{equation}
Here $j$ labels lattice sites, $x_j$ is the coordinate along the imposed gradient, and the model parameters follow the definitions in the main text.
Quantum fluctuations enter through stochastic sampling of the initial Wigner distribution, and observables are obtained from ensemble averages over many trajectories.
\begin{figure*}[t]
  \centering
  \includegraphics[width=0.8\textwidth]{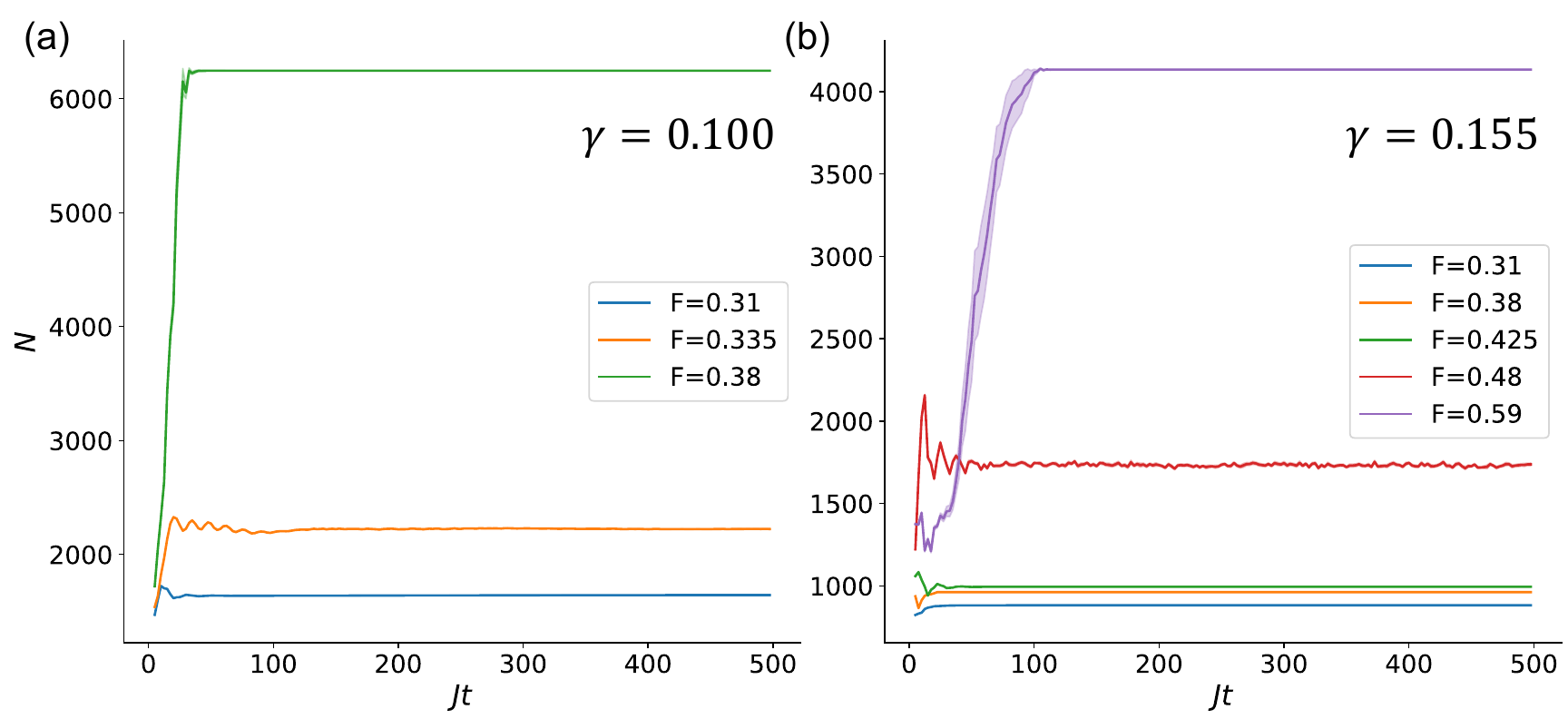}
  \caption{TWA ensemble for the total occupation. (a) $\gamma=0.100$; (b) $\gamma=0.155$. Solid lines show the ensemble mean; shaded bands indicate the sample standard deviation across $M=50$ trajectories (the standard error of the mean scales as $1/\sqrt{M}$). The visible band at $F=0.59$ in panel (b) reflects long transients and competition among long-lived dynamical responses when a stationary mean-field solution is absent or unstable at these parameters. Whenever a stable stationary response exists, trajectories collapse rapidly and the dispersion falls below the plotting linewidth. Simulation parameters match those in the main text.}
  \label{fig:SM_TWA}
\end{figure*}

\subsection{A. Initialization and observables}
For each trajectory, the initial condition is prepared by adding a small coherent seed together with vacuum Wigner noise:
\begin{equation}
\alpha_j(0)=\alpha^{\rm box}_j+\xi^{(0)}_j,
\label{eq:S_TWA_init}
\end{equation}
where $\alpha^{\rm box}_j$ is sampled independently on each lattice site from a uniform distribution over the rectangle
$\{x+iy \mid x\in[0,0.5],\, y\in[-0.5,0]\}$ in the complex plane. The vacuum Wigner noise $\xi^{(0)}_j$ is a complex Gaussian random variable with mean $\langle \xi^{(0)}_j\rangle=0$ and variance $1/4$ per quadrature.

\subsection{B. Numerical implementation}
For our simulations, we sample the initial Wigner noise at $t=0$ as described above and then propagate the fields according to Eq.~\!\eqref{eq:S_TWA_eom} using a fourth-order Runge-Kutta integrator with the time step $dt=0.01$. This protocol is well suited to the large-occupation regime and the extensively averaged observables emphasized here, thereby justifying the IV-TWA treatment.

Figure~\!\ref{fig:SM_TWA} illustrates trajectory-to-trajectory dispersion within an IV-TWA ensemble of $M=50$ trajectories; the shaded bands therefore quantify ensemble variability. For $\gamma=0.100$, as shown in Fig.~\!\ref{fig:SM_TWA}(a), different realizations relax rapidly toward the same stationary response, resulting in a dispersion that is at or below the plotting linewidth. For $\gamma=0.155$ (in Fig.~\!\ref{fig:SM_TWA}(b)), the dispersion becomes visible at large $F$, reflecting prolonged transients and the competition among long-lived dynamical responses when the stationary mean-field solution is absent or unstable. In the parameter regimes classified as chaotic, the extensive occupation remains strongly self-averaging over the lattice, and its ensemble spread stays comparatively small.

\subsection{C. Scope and limitations}
TWA neglects higher-order terms beyond leading order in the Wigner expansion, and its accuracy decreases when the occupation per mode is small. Within the regimes explored here (high-density fronts and large-scale collective dynamics), IV-TWA reproduces the mean-field structures (pinned states and fronts) while capturing fluctuation-induced effects (destabilization and switching), thereby providing an appropriate bridge between semiclassical and quantum descriptions in driven-dissipative Bose-Hubbard lattices.

\section{II. One-dimensional reduction and front pinning condition}

We start from the two-dimensional gGPE,
\begin{equation}
i\frac{d\alpha_{x,y}}{dt}
=\Big[\Delta\omega+Fx-\tfrac{i}{2}\big(\kappa_0+\gamma x\big)\Big]\alpha_{x,y}
+U|\alpha_{x,y}|^2\alpha_{x,y}
-J\big(\alpha_{x+1,y}+\alpha_{x-1,y}+\alpha_{x,y+1}+\alpha_{x,y-1}\big)
+\Omega ,
\label{eq:2D-SM}
\end{equation}
where $\alpha_{x,y}$ is the coherent amplitude on site $(x,y)$, $F$ is the Stark tilt that sets the real detuning ramp, and $\kappa(x)=\kappa_0+\gamma x$ is the spatially varying loss rate. It is convenient to rewrite the hopping term using the discrete identity
\begin{equation}
\sum_{\langle x',y'\rangle}\alpha_{x',y'}
=4\alpha_{x,y}+\Delta_d\alpha_{x,y},
\end{equation}
where $\Delta_d\alpha_{x,y}
=\alpha_{x+1,y}+\alpha_{x-1,y}+\alpha_{x,y+1}+\alpha_{x,y-1}-4\alpha_{x,y}$. This gives
\begin{equation}
-J\!\!\sum_{\langle x',y'\rangle}\!\alpha_{x',y'}
=-4J\alpha_{x,y}-J\Delta_d\alpha_{x,y}.
\label{eq:hop-decomp}
\end{equation}
The first contribution is a fixed coordination shift on the square lattice, while the second is a discrete Laplacian. In the pinned-front regime of interest, numerical simulations show that the front is nearly uniform along the transverse ($y$) direction. We therefore set $\alpha_{x,y\pm1}\approx\alpha_{x,y}$ in the front region, so that
\begin{equation}
\Delta_d\alpha_{x,y}\approx \alpha_{x+1,y}+\alpha_{x-1,y}-2\alpha_{x,y}.
\end{equation}
Imposing the steady-state condition $\partial_t\alpha=0$ then gives the reduced one-dimensional equation
\begin{equation}
-J\big(\alpha_{x+1}+\alpha_{x-1}-2\alpha_x\big)
+\Big[\Delta_{1\mathrm D}(x)-\tfrac{i}{2}\kappa(x)\Big]\alpha_x
+U|\alpha_x|^2\alpha_x+\Omega=0,
\label{eq:1Dsteady-SM}
\end{equation}
with $\Delta_{1\mathrm D}(x)=\Delta\omega+Fx-4J$. The shift $-4J$ is fixed by the square-lattice coordination number.

\subsection{A. Local-density approximation and resonant-slice pinning}
Neglecting the term $-J(\alpha_{x+1}+\alpha_{x-1}-2\alpha_x)$ in a local-density approximation (LDA) yields the local input-output relation
\begin{equation}
\frac{\Omega^2}{N(x)}=\big[\Delta_{1\mathrm D}(x)-U N(x)\big]^2+\frac{1}{4}\kappa^2(x),
\quad N(x)=|\alpha_x|^2,
\label{eq:cubic-SM}
\end{equation}
and motivates the effective detuning $\Delta_{\mathrm{eff}}(x)=\Delta_{1\mathrm D}(x)-U N(x)$. The pinned front is centered at the resonant slice $\Delta_{\mathrm{eff}}(x_s)=0$, where
\begin{equation}
N_s=\frac{4\Omega^2}{\kappa^2(x_s)},
\quad
\Delta\omega+F x_s-4J=\frac{4U\Omega^2}{\kappa^2(x_s)}.
\label{eq:pin-SM}
\end{equation}
This admits a unique solution for $x_s$.

\subsection{B. Front width}
To estimate the width of the pinned front, we retain the discrete-Laplacian contribution in Eq.~\!\eqref{eq:1Dsteady-SM} and focus on the vicinity of the resonant slice $x_s$ defined by $\Delta_{\mathrm{eff}}(x_s)=0$. Near the resonant position $x_s$, we linearize
\begin{equation}
\Delta_{\mathrm{eff}}(x)\equiv \Delta_{1\mathrm D}(x)-U N(x)\;\approx\; S\cdot(x-x_s),
\qquad 
S=\frac{\dd\Delta_{\mathrm{eff}}}{\dd x}\Big|_{x_s}
=F-U\frac{\dd N}{\dd x}\Big|_{x_s}.
\label{eq:width-SM}
\end{equation}
Here $S$ is the local slope of $\Delta_{\mathrm{eff}}(x)$ at the front. A larger $S$ therefore implies a sharper front. Balancing the discrete-Laplacian contribution against the linearized detuning in Eq.~\!\eqref{eq:1Dsteady-SM} yields an Airy-type envelope equation near $x_s$, and the corresponding length scale follows from the scaling estimate $J/w^2 \sim S w$, giving
\begin{equation}
w\sim \left(\frac{J}{S}\right)^{1/3}.
\end{equation}
To obtain $S$ explicitly within the LDA, we differentiate Eq.~\!\eqref{eq:cubic-SM} at $\Delta_{\mathrm{eff}}=0$ and find
\begin{equation}
\frac{\kappa^2(x_s)}{4}\frac{\dd N}{\dd x}\Big|_{x_s}
+ \frac{1}{2}\gamma N_s\kappa(x_s)=0
\quad\Rightarrow\quad
\frac{\dd N}{\dd x}\Big|_{x_s}=-\frac{2\gamma}{\kappa(x_s)}N_s,
\label{eq:dNdx-SM}
\end{equation}
and therefore
\begin{equation}
S \approx F+\frac{2\gamma U}{\kappa(x_s)}N_s
=F+\frac{8\gamma U\Omega^2}{\kappa^3(x_s)}.
\label{eq:S-SM}
\end{equation}

\subsection{C. Downstream tail}
For $x>x_s$ sufficiently far from resonance, where $U N(x)\ll|\Delta_{1\mathrm D}|$, Eq.~\eqref{eq:cubic-SM} reduces to
\begin{equation}
N(x)\approx \frac{4\Omega^2}{4\Delta_{1\mathrm D}^2(x)+\kappa^2(x)}.
\end{equation}
A convenient local attenuation length is then
\begin{equation}
\ell_+(x)=\left|\frac{\dd\ln N(x)}{\dd x}\right|^{-1}
=
\left|
\frac{8F\Delta_{1\mathrm D}(x)+2\gamma\kappa(x)}
{4\Delta_{1\mathrm D}^2(x)+\kappa^2(x)}
\right|^{-1}.
\end{equation}
In the parameter regimes of interest, $\ell_+(x)$ is typically much larger than $w$, separating the sharp front width from the smoother downstream decay.

\subsection{D. Discrete-Laplacian correction to the pinning condition}
For quantitative comparison with simulations while keeping the lattice shift $-4J$ fixed, lattice discreteness can be incorporated as an effective detuning correction associated with the discrete-Laplacian contribution in Eq.~\!\eqref{eq:1Dsteady-SM}. We define
\begin{equation}
\delta\Delta_{\mathrm{lap}}(x) = J\,\mathrm{Re}\left[\frac{\big(\alpha_{x+1}+\alpha_{x-1}-2\alpha_x\big)\alpha_x^\ast}{|\alpha_x|^2}\right],
\label{eq:lap-correction}
\end{equation}
which depends only on the local steady-state profile. The refined pinning condition becomes
\begin{equation}
\Delta\omega+F x_s-4J+\delta\Delta_{\mathrm{lap}}(x_s)
= U N_s.
\label{eq:pin-refined}
\end{equation}
When the discrete-Laplacian correction is negligible, Eq.~\!\eqref{eq:pin-refined} reduces to Eq.~\!\eqref{eq:pin-SM}. For broad fronts one expects
\begin{equation}
\mathrm{Re}\!\left[\frac{\alpha_{x+1}+\alpha_{x-1}-2\alpha_x}{\alpha_x}\right]
\sim \mathcal{O}\!\left(\frac{a^2}{w^2}\right),
\end{equation}
where $a$ is the lattice spacing, so that $\delta\Delta_{\mathrm{lap}}$ vanishes as $w$ increases.

\section{III. Transverse instability and pattern locking}
As established above, the pinning position $x_s$ is fixed by a self-consistent resonance condition, while the intrinsic front width $w$ follows from a balance between curvature and the local detuning gradient. We now incorporate slow modulations of the front along the transverse direction.

Transverse modulations are assumed to be long-wavelength on the scale of the lattice, so that $y$ may be treated as a continuous coordinate when forming gradients, while the underlying microscopic dynamics remains that of $\alpha_{x,y}(t)$. Under this separation of scales, we introduce the modulated-front ansatz
\begin{equation}
  \alpha_{x,y}(t)\approx \alpha_0\big(x-X(y,t)\big),
\end{equation}
where $\alpha_0(x)$ denotes the stationary one-dimensional pinned profile centered at $x_s$, and $X(y,t)$ is a slowly varying front displacement.

Substituting the ansatz into the equations of motion and performing a systematic long-wavelength expansion in transverse gradients yields a solvability condition associated with the front-displacement mode $\partial_x\alpha_0$. Enforcing this condition by projection onto the corresponding adjoint mode gives the effective amplitude equation
\begin{equation}
  \partial_t X = -\mu - D\partial_y^2 X - K\partial_y^4 X + \Lambda(\partial_y X)^2 + \cdots ,
  \label{eq:SM_amp}
\end{equation}
where $\mu$ is an effective bias term. For an exactly pinned interior front, the net drift vanishes and $\mu=0$. The coefficients $D$, $K$, and $\Lambda$ are determined by overlap integrals involving the adjoint mode and the operators generated by the gradient expansion. Let $\phi$ solve the adjoint eigenproblem associated with the linearization about $\alpha_0$ and impose the normalization $\langle \phi,\partial_x\alpha_0\rangle=1$. One then obtains
\begin{equation}
  D=\langle \phi,\mathcal{P}_2[\alpha_0]\rangle,\quad
  K=\langle \phi,\mathcal{P}_4[\alpha_0]\rangle,\quad
  \Lambda=\langle \phi,\mathcal{N}[\alpha_0]\rangle,
  \label{eq:SM_proj}
\end{equation}
where $\mathcal{P}_{2}$ and $\mathcal{P}_{4}$ collect the contributions that are second and fourth order in $\partial_y$, and $\mathcal{N}$ denotes the leading quadratic nonlinearity generated by the modulation. In the parameter regime considered here, $K>0$, which provides short-wavelength regularization consistent with a finite front width. The dependence of $(D,K,\Lambda)$ on microscopic control parameters enters through the resonant-slice slope $S$ and the width scaling $w\propto(J/S)^{1/3}$. This implies $K\propto J w^2$ decreases as $S$ increases. The coefficient $D$ also decreases with increasing $S$ and can change sign, thereby triggering a transverse instability.

Linearizing Eq.~\eqref{eq:SM_amp} about a flat front yields the dispersion relation
\begin{equation}
  \sigma(k_y) = -D k_y^2 - K k_y^4,\quad k_y=\frac{2\pi m}{L_y},
  \label{eq:SM_disp}
\end{equation}
where $k_y$ denotes the transverse wavevector, and $m\in\mathbb{Z}$ labels transverse Fourier modes on a strip of width $L_y$ with periodic boundary conditions. When $D<0$ and $K>0$, an instability band emerges for
\begin{equation}
  0<k_y<\sqrt{\frac{|D|}{K}},
  \quad
  k_{y,\max}=\sqrt{\frac{|D|}{2K}},
  \label{eq:SM_kmax}
\end{equation}
with maximal growth rate attained at $k_{y,\max}$. The discreteness of allowed $k_y$ on a finite strip leads to mode selection and locking onto the Fourier harmonic closest to $k_{y,\max}$. As $|D|$ increases, the unstable band broadens. Nonlinear mode coupling driven by the quadratic term in Eq.~\eqref{eq:SM_amp} then promotes a crossover to a Kuramoto--Sivashinsky-type regime~\cite{PhysRevA.90.023615, PhysRevB.91.045301, RevModPhys.97.025004}, characterized by spatiotemporal chaos and broadband fluctuations in extensive observables, including the total occupation $N(t)$ defined in the main text.

For sufficiently large Stark tilt $F$, the selected interior front ceases to exist and the high-density region becomes localized near the minimum-loss side. This transition can be rationalized by two complementary considerations. First, a geometric criterion is obtained by comparing the pinned position $x_s(F,\gamma)$ with the intrinsic width $w(F,\gamma)$. Once $x_s(F,\gamma)$ approaches the minimum-loss side on the scale of $w(F,\gamma)$, the distinction between an interior front and a minimum-loss-localized layer is no longer meaningful, leading to the condition
\begin{equation}
  x_s(F,\gamma) \lesssim c\,w(F,\gamma),
  \label{eq:SM_geom}
\end{equation}
where $c$ is a nonuniversal constant of order unity. Second, the combination of transverse activity and the dissipation gradient produces a systematic renormalization of the drift term in Eq.~\eqref{eq:SM_amp}. This effect can be captured by an effective bias $\mu_{\mathrm{eff}}$ that becomes negative at large $F$, corresponding to net motion toward the minimum-loss side. In numerics, both effects are observed: at large $F$ the global density increases rapidly and subsequently saturates in a stable minimum-loss-localized layer.

\section{IV. Dynamical phase-transition pathways}
\label{sec:SM_transition_pathways}

The dynamical phase diagram in the main text separates three long-time responses. Phase I is an interior pinned-front state with $x_s>0$, including both transversely uniform fronts and finite-size pattern-locked fronts. Phase II is the chaotic continuation of the same interior front, where nonlinear coupling among unstable transverse modes sustains broadband spatiotemporal fluctuations. Phase III is the minimum-loss-localized state, where the high-density region is localized at the minimum-loss side ($x=0$).

The transition pathway is selected by the ordering of two boundary values along a given parameter cut. 

The first, denoted $F_{\mathrm{ch}}(\gamma)$, marks the onset of sustained chaotic dynamics in the phase diagram. The linear mechanism is controlled by Eq.~\eqref{eq:SM_disp}: a flat front is linearly stable for $D>0$, whereas $D<0$ with $K>0$ opens the unstable band in Eq.~\eqref{eq:SM_kmax}. On a finite strip this instability may first saturate into a pattern-locked front, which we still classify as Phase I. The chaotic regime is reached only when increasing $|D|$ broadens the unstable band enough that nonlinear mode coupling in Eq.~\eqref{eq:SM_amp} generates persistent broadband dynamics.  The second boundary value, denoted $F_{\mathrm{loc}}(\gamma)$, marks minimum-loss localization. As $F$ increases, the resonant slice $x_s(F,\gamma)$ moves toward the minimum-loss side. The interior-front description ceases to be distinct from a minimum-loss-localized layer once the front position becomes comparable to its width, namely
\begin{equation}
  x_s(F,\gamma) \lesssim c\,w(F,\gamma).
  \label{eq:SM_Floc}
\end{equation}
Beyond this localization boundary the dynamics relax to Phase III. The two possible orderings are therefore

\begin{equation}
\begin{array}{ll}
F_{\mathrm{loc}}<F_{\mathrm{ch}}: & \mathrm{I}\to\mathrm{III},\\
F_{\mathrm{ch}}<F_{\mathrm{loc}}: & \mathrm{I}\to\mathrm{II}\to\mathrm{III}.
\end{array}
\label{eq:SM_pathways}
\end{equation}

\section{V. Staircase in the Total Particle Number versus Stark Tilt}

A characteristic staircase structure in the total particle number
\[
N(F)=\sum_{j}|\alpha_j|^2
\]
emerges when the Stark tilt $F$ is increased at fixed dissipation gradient $\gamma$. This effect can be explained within the resonant-slice framework developed in the main text. The interior front (bright band) is centered at $x_s(F,\gamma)$, determined by the self-consistent resonance condition
\begin{equation}
\Delta\omega + F x_s - 4J = \frac{4U\Omega^2}{\kappa^2(x_s)} 
=:R(x_s),
\label{eq:SM_step_pin}
\end{equation}
where $\kappa(x)=\kappa_0+\gamma x$. On a discrete lattice, the front additionally experiences a Peierls-Nabarro (PN) locking potential~\cite{PhysRevA.69.053604, Jenkinson:2017aa} that favors alignment with integer $x$ positions. To capture this discreteness, one may write the normal form
\begin{equation}
\Delta\omega + F x_s - 4J = R(x_s) + \varepsilon_{\mathrm{PN}} \sin(2\pi x_s),
\label{eq:SM_step_PN}
\end{equation}
where $\varepsilon_{\mathrm{PN}}$ is an exponentially small amplitude controlled by the intrinsic width of the front (see below). For a given $F$, Eq.~\eqref{eq:SM_step_PN} admits solution branches with $x_s$ localized near integer positions. Each branch persists until a saddle-node bifurcation (fold) is reached, at which point the front depins to the neighboring integer $x$ position. Differentiation of Eq.~\eqref{eq:SM_step_PN} with respect to $x_s$ gives the fold condition
\begin{equation}
F - R'(x_s) - 2\pi \varepsilon_{\mathrm{PN}}\cos(2\pi x_s)=0,
\quad
R'(x)=-\frac{8U\Omega^2\gamma}{\big(\kappa_0+\gamma x\big)^3}<0.
\label{eq:SM_step_fold}
\end{equation}
As $F$ is increased, the sequence of folds produces a stick-slip (pin-depin) motion of the resonant slice, with $x_s$ advancing approximately one lattice spacing per event. This mechanism directly yields the observed staircase in $N(F)$.

The sensitivity of the resonant position to changes in $F$ can be obtained in the absence of PN locking by implicit differentiation of Eq.~\eqref{eq:SM_step_pin}:
\begin{equation}
\frac{dx_s}{dF}=-\frac{x_s}{F-R'(x_s)}.
\label{eq:SM_step_sensitivity}
\end{equation}
Because $R'(x_s)<0$, increasing $\gamma$ enhances $|R'(x_s)|$ and thereby reduces $|dx_s/dF|$. This weakens the drift of the front with $F$, decreases the frequency of depinning events, and smooths the staircase profile.

The PN amplitude $\varepsilon_{\mathrm{PN}}$ is governed by the front width $w$. From the main text, the slope of the effective detuning is
\begin{equation}
S=\frac{d\Delta_{\mathrm{eff}}}{dx}\Big|_{x_s}
\approx F + \frac{8U\gamma\Omega^2}{\kappa(x_s)^3},
\quad
w\propto\left(\frac{J}{S}\right)^{1/3}.
\label{eq:SM_step_width}
\end{equation}
Thus $S$ increases and $w$ decreases as either $F$ or $\gamma$ increases. The PN amplitude follows the exponential law
\begin{equation}
\varepsilon_{\mathrm{PN}} \propto J \exp\!\left(-\eta_{\mathrm{PN}} \frac{w}{a}\right),
\label{eq:SM_step_epsilon}
\end{equation}
where $a$ is the lattice spacing and $\eta_{\mathrm{PN}}=\mathcal{O}(1)$ is a dimensionless discreteness constant. A larger $\gamma$ (or larger $F$) therefore increases $\varepsilon_{\mathrm{PN}}$, which stabilizes each pinned branch over a wider interval of $F$ through Eq.~\!\eqref{eq:SM_step_fold}. Combined with Eq.~\!\eqref{eq:SM_step_sensitivity}, this explains why the staircase becomes less frequent and smoother as $\gamma$ increases.

The magnitude of each step in $N$ is set by removing one upstream $x$ slice when the front advances by one lattice site. Approximating the upstream density as slowly varying, one finds
\begin{equation}
\Delta N \approx L_y n_{\mathrm{up}}(x_s),
\quad
n_{\mathrm{up}}(x)\approx \frac{4\Omega^2}{\kappa^2(x)}\quad \text{near resonance},
\label{eq:SM_step_jump}
\end{equation}
where $L_y$ is the system size in the transverse direction. Since $\kappa(x)$ increases with $\gamma$, the upstream density decreases and the step height $\Delta N$ is reduced at larger $\gamma$. This trend accounts for the systematically smaller staircase amplitudes observed in simulations as the dissipation gradient is increased.
 
In parameter regions where the interior front simultaneously develops transverse modulations or evolves into a minimum-loss-localized layer, the discrete stick-slip dynamics coexist with switching between long-lived responses. In such cases, the staircase is superimposed with larger apparent steps or spike-like features in $N(F)$. Away from these regions, however, the staircase structure is dominated by the Peierls-Nabarro depinning mechanism encapsulated by Eqs.\!~\eqref{eq:SM_step_PN}-\eqref{eq:SM_step_jump}.

\section{VI. Definition of the Generalized Imbalance}
In studies of many-body localization, imbalance diagnostics commonly quantify memory of an initially prepared density pattern~\cite{szdc-61nl,Mak:2024aa,Schreiber2015,PhysRevLett.119.260401}. For a density-wave initial state, the conventional imbalance is the normalized difference between the particle numbers on even and odd sites,
\begin{equation}
\mathcal{I}(t) = \frac{N_{\mathrm{even}}(t)-N_{\mathrm{odd}}(t)}{N_{\mathrm{even}}(t)+N_{\mathrm{odd}}(t)} ,
\end{equation}
which directly probes the relaxation of an initially prepared density-wave pattern. For a perfect density-wave initial state, one finds $\mathcal{I}(0)=1$. In a thermal phase, $\mathcal{I}(t)$ decays to zero at long times, while in a localized phase it remains finite. The above definition can be understood more generally as the overlap of the density distribution at time $t$ with a fixed reference pattern $m_j$. In the case of the density wave, $m_j=+1$ on even sites and $m_j=-1$ on odd sites. The imbalance thus quantifies the degree to which the system retains memory of a specific initial configuration.

In our driven-dissipative setting, however, the initial state is not a fixed density wave but is instead drawn randomly as shown in Eq.~\!\eqref{eq:S_TWA_init}. We therefore use the term generalized imbalance operationally, replacing the fixed even-odd reference pattern with stochastic density profiles drawn from the initialization ensemble. For a paired trajectory, the corresponding density-overlap diagnostic would be
\begin{equation}
\mathcal{I}(t) \;=\; \frac{\sum_j N_j(t)\, N_j(0)}{\sum_j N_j^2(0)}, \quad N_j(t)=|\alpha_j(t)|^2 .
\end{equation}
By construction, the denominator sets $\mathcal{I}(0)=1$ for each trajectory. At later times the diagnostic measures the overlap of the evolved density with the reference profile after normalization by the initial density self-overlap. For the phase-map diagnostic used in the main text, each stochastic trajectory is evolved from its own initial state and the paired density-overlap diagnostic is averaged over the ensemble. The generalized imbalance that we employ is therefore
\begin{equation}
\mathcal{I}(t) = \frac{1}{M}\sum_{k=1}^{M}
\frac{\sum_j N_j^{k}(t)N_j^{k}(0)}{\sum_j \left[N_j^{k}(0)\right]^2},
\end{equation}
where $M=50$ is the number of stochastic trajectories and $N_j^{k}(t)$ is the density on site $j$ at time $t$ in trajectory $k$. This provides a front-independent density-overlap diagnostic for the present setting.

\begin{figure}[htbp]
  \centering
  \includegraphics[width=0.7\linewidth]{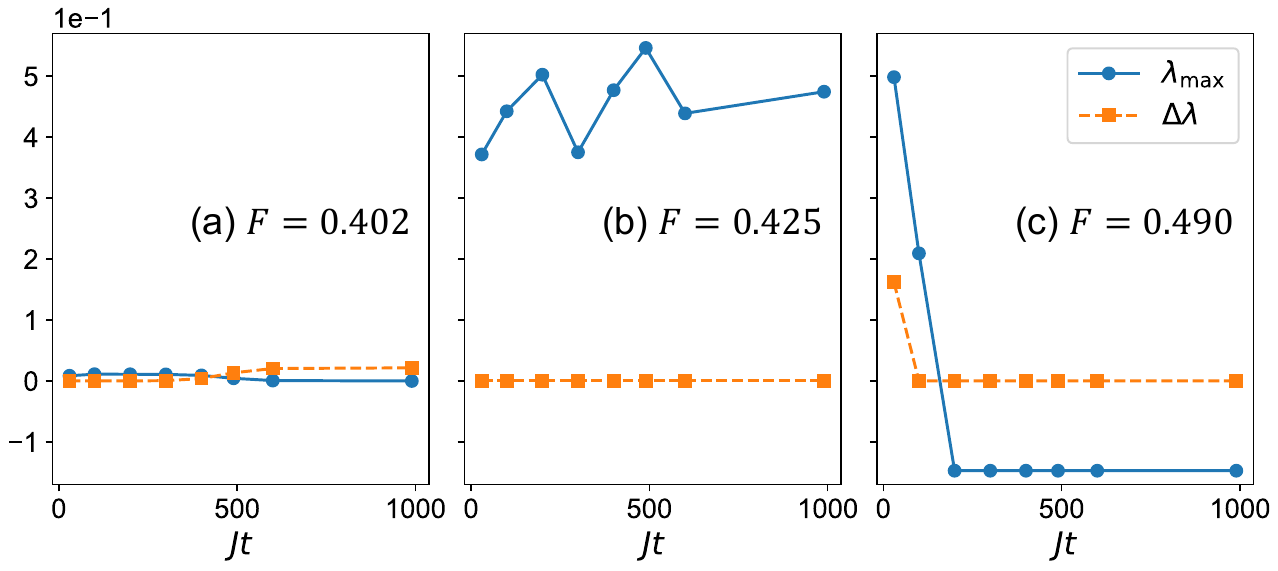}
  \caption{Time-resolved Bogoliubov-de Gennes (BdG) linear-stability diagnostic. At each plotted time $Jt$, the driven-dissipative gGPE is linearized around the instantaneous numerical state. The leading real part $\lambda_{\text{max}}$ indicates whether perturbations grow locally under the corresponding linearized dynamics; $\Delta\lambda$ denotes the difference between the largest and second-largest real parts of the BdG eigenvalues.}
\label{fig:lambda}
\end{figure}

\section{VII. Time-Resolved Bogoliubov-de Gennes Linear-Stability Diagnostic}
As an auxiliary check, we linearize the driven-dissipative gGPE along the simulated time evolution. At each sampled time, perturbations $\delta\alpha_j$ around the instantaneous numerical state obey $\frac{\mathrm{d}}{\mathrm{d}t}(\delta\alpha_j, \delta\alpha^*_j)^{\top} = \mathcal{L}(t)(\delta\alpha_j, \delta\alpha^*_j)^{\top}$,
where $\mathcal{L}(t)$ is the non-Hermitian BdG generator~\cite{PhysRevA.110.053305, PhysRevA.105.023319, PhysRevA.98.053605}. We order the BdG eigenvalues by their real parts, $\mathrm{Re}\,\lambda_0(t)\ge \mathrm{Re}\,\lambda_1(t)\ge \cdots$, and plot $\lambda_{\text{max}}(t)=\mathrm{Re}\,\lambda_0(t)$ together with $\Delta\lambda(t)=\mathrm{Re}\,\lambda_0(t)-\mathrm{Re}\,\lambda_1(t)$. Thus $\Delta\lambda$ measures the real-part separation of the leading BdG growth rate.

A positive $\lambda_{\text{max}}(t)$ indicates local linear growth of perturbations around the instantaneous state and is consistent with evolutions that develop transverse patterning or spatiotemporal chaos. Values near zero indicate that no strong positive instantaneous BdG instability is resolved at that time.

\providecommand{\noopsort}[1]{}\providecommand{\singleletter}[1]{#1}%